\documentclass[lettersize,journal]{IEEEtran}
\usepackage{cite}
\usepackage{amsmath,amssymb,amsfonts}
\usepackage{algorithmic}
\usepackage{graphicx}
\usepackage{textcomp}
\usepackage{xcolor}

\usepackage{flushend}

  \renewenvironment{thebibliography}[1]{%
    \begin{oldthebibliography}{#1}%
      \setlength{\parskip}{-0.00ex}%
      \setlength{\itemsep}{-0.0ex}%
  }%
  {%
    \end{oldthebibliography}%
  }

\newcommand{\ignore}[1]{ }

\usepackage{url}
\usepackage{colortbl}
\usepackage{hhline}
\usepackage{relsize}
\usepackage[hidelinks,colorlinks=true,linkcolor=blue,citecolor=cyan,urlcolor=black]{hyperref}
\usepackage{pifont}
\usepackage{float}

\usepackage[normalem]{ulem}

\usepackage{balance}

\usepackage{marvosym}
\usepackage{ragged2e}
\usepackage{multirow}
\usepackage{vcell}
\usepackage{comment}

\usepackage{courier}
\usepackage{balance}

\usepackage{hhline}
\usepackage{arydshln}
\usepackage{textcomp}

 \usepackage{booktabs}

\usepackage{diagbox}

\newcommand{\red}{}

\newcommand{\blue}{}
\renewcommand{\red}{\textcolor{red}}

\renewcommand{\blue}{\textcolor{blue}}

\usepackage{soul}

\usepackage[hang,flushmargin]{footmisc} 

\usepackage{tcolorbox} 

\usepackage{rotating}

\usepackage{ifthen}
\newboolean{showcomments}
\setboolean{showcomments}{false}
\ifthenelse{\boolean{showcomments}}
{ \newcommand{\mynote}[3]{
     \fbox{\bfseries\sffamily\scriptsize#1}   {\small$\blacktriangleright$\textsf{\emph{\color{#3}{#2}}}$\blacktriangleleft$}}}
{ \newcommand{\mynote}[3]{}}

\usepackage{tikz,xcolor}

\definecolor{lime}{HTML}{A6CE39}
\DeclareRobustCommand{\orcidicon}{%
	\begin{tikzpicture}
	\draw[lime, fill=lime] (0,0) 
	circle [radius=0.14] 
	node[white] {{\fontfamily{qag}\selectfont \tiny ID}};
	\draw[white, fill=white] (-0.0625,0.095) 
	circle [radius=0.007];
	\end{tikzpicture}
	\hspace{-2mm}
}

\foreach \x in {A, ..., Z}{%
	\expandafter\xdef\csname orcid\x\endcsname{\noexpand\href{https://orcid.org/\csname orcidauthor\x\endcsname}{\noexpand\orcidicon}}
}

\newcommand{\Design}{\textbf{\texttt{TranSC}}}

\begin{document}


\title{
TranSC: Hardware-Aware Design of Transcendental Functions Using Stochastic Logic
}

\author{IEEE Publication Technology,~\IEEEmembership{Staff,~IEEE,}
\thanks{This paper was produced by the IEEE Publication Technology Group. They are in Piscataway, NJ.}
\thanks{Manuscript received April 19, 2021; revised August 16, 2021.}}



\author{
        
        Mehran~Moghadam\textsuperscript{\orcidA{}},~\IEEEmembership{Graduate Member,~IEEE}, Sercan~Aygun\textsuperscript{\orcidB{}},~\IEEEmembership{Senior Member,~IEEE}, and \\ M.~Hassan~Najafi\textsuperscript{\orcidE{}},~\IEEEmembership{Senior Member,~IEEE
        \vspace{-2em}
        }


\thanks{
\footnotesize{
   This work is supported in part by National Science Foundation (NSF) under grants \#2019511, \#2339701, NASA grant 80NSSC25C0335, and generous gifts from NVIDIA.
   Mehran Moghadam and M. Hassan Najafi are with the Electrical, Computer, and Systems Engineering Department, Case Western Reserve University, Cleveland, OH, 
   USA. 
   E-mail:\{moghadam, najafi\}@case.edu. Sercan Aygun is with the School of Computing and Informatics, University of Louisiana at Lafayette, Lafayette, LA, USA. E-mail: sercan.aygun@louisiana.edu. 
A preliminary version of this paper appeared as~\cite{TriSC_DAC24}.
   }
}
}

\maketitle


\begin{abstract}
The hardware-friendly implementation of transcendental 
functions remains a longstanding challenge in design automation. 
These functions, which cannot be expressed as finite combinations of algebraic operations, pose significant complexity in digital circuit design.
This study introduces a novel approach, \Design, that utilizes stochastic computing (SC) for lightweight yet accurate implementation of transcendental functions. Building on established SC techniques, 
our method explores alternative random sources—specifically, quasi-random Van der Corput low-discrepancy (LD) sequences—instead of conventional pseudo-randomness. This shift enhances both the accuracy and efficiency of SC-based computations. 
We validate our approach through extensive experiments on various function types, including trigonometric, hyperbolic, and activation functions. 
The proposed design approach significantly reduces MSE 
by up to 98\% 
compared to the state-of-the-art solutions while reducing hardware area, power consumption, and energy usage by 33\%, 72\%, and 64\%, respectively.
\end{abstract}
\vspace{-0.5em}


\vspace{-1.5em}
\section{Introduction}
\label{introduction}
\IEEEPARstart{T}{he} rapid growth of deep neural networks has placed unprecedented demands on computing hardware, particularly for activation functions that introduce non-linearity. Functions such as sigmoid, hyperbolic tangent, and their underlying exponential and logarithmic forms are especially costly when implemented using conventional binary computing. 
Likewise, trigonometric functions such as sine and cosine are indispensable for analyzing periodic behavior in digital signal processing (DSP), 
with widespread applications in motor control, communications, and noise filtering. These domains often operate under strict real-time and low-power constraints. 
Implementing transcendental functions in software on general-purpose processors causes significant computational delays, ultimately limiting throughput and responsiveness. 
To overcome this bottleneck, the design of low-latency, cost-efficient hardware accelerators has become a critical objective for embedded DSP systems~\cite{Dalloo_2024}. 

Transcendental functions are equally vital in computer graphics and robotics,  where they play a central role in coordinate transformations, lighting calculations, and robotic kinematics. 
These operations frequently depend on sine and cosine computations, and their accuracy directly impacts the final output, whether in the visual quality of a rendered image or the positioning precision of a robotic manipulator.
The importance of these functions is underscored by the inclusion of dedicated Special Function Units (SFUs) in modern GPUs to accelerate their evaluation~\cite{Ang_SFU2016}. However, in resource-constrained platforms such as mobile robots, autonomous drones, or AR/VR headsets, the power and area requirements of SFUs can be prohibitive.  This limitation has created a strong demand for alternative hardware solutions that are both ultra-lightweight and energy-efficient, while still providing sufficient accuracy for performance-critical applications.
\color{black}

\begin{figure}[t]
  \centering
\includegraphics[width=\linewidth]{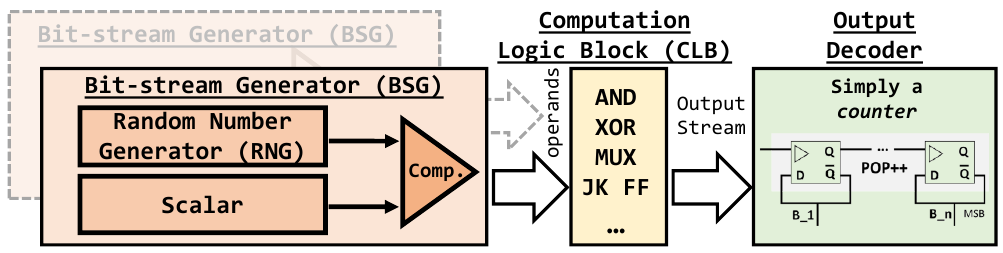}
  \vspace{-1.5em}
  \caption{A general 
  SC system architecture with BSG, CLB, and output decoder 
  that converts the output bit-stream to standard radix representation.}
  \label{figure1}
  \vspace{-1.5em}
\end{figure}

Stochastic computing (SC) is a re-emerging computing paradigm that enables ultra-low-cost implementations for complex arithmetic operations~\cite{10.1145/2465787.2465794, 10460194}. 
SC offers robust data processing with significantly reduced hardware cost and power consumption, making it attractive for resource-constrained systems. An SC system typically comprises three core modules:  a \textit{bit-stream generator (BSG)}, a \textit{computation 
logic block (CLB)}, and an \textit{output decoder} \cite{9165977}. In SC systems, 
scalar values are converted into 
random binary sequences, known as bit-streams, where all bits have equal weight. 
The quality of randomness in these bit-streams (governed by the BSG, often implemented with pseudo-random number generators (RNGs)) plays a critical role in determining system accuracy.
Fig.~\ref{figure1} illustrates a 
general SC system 
consisting of the BSG, CLB, and output decoder. The CLB unit processes the bit-streams using simple logic gates (e.g., \texttt{AND}, \texttt{OR}, \texttt{XOR}) or lightweight sequential circuits such as 
flip-flops (\texttt{FF}s). 
The decoder converts the output bit-stream back to standard binary radix format, typically 
by counting the number of ones in the bit-stream~\cite{bsbnn}. 
This architecture allows SC to replace costly arithmetic circuits with basic logic components, offering a highly efficient alternative for implementing complex mathematical functions.

Numerous arithmetic blocks have been developed within SC frameworks~\cite{electronics8060720, 9741299, 9045292}. In particular, considerable research has focused on extending SC to transcendental and algebraic functions~\cite{10328839,Trig-Parhi,9444648,10527390,9843901}, often by employing intermediary mathematical tools, such as Bernstein polynomials and Taylor or Maclaurin series expansions.
While prior efforts have primarily concentrated on designing low-cost CLB units~\cite{10.1145/3007648}, this work shifts attention to the BSG as a key lever for improving accuracy and reducing hardware overhead and processing latency.
Specifically, we exploit lightweight low-discrepancy (LD) sequence generators for the BSG unit to realize highly efficient SC designs.
LD sequences are quasi-random number generators that distribute values more uniformly than pseudo-random sources, enabling faster convergence and reduced error in SC-based designs~\cite{8327916}. The key contributions of this study are as follows: 
\\
\ding{182} Design exploration of SC-based transcendental 
functions using lightweight \textit{Quasi-Random} BSG units. 
\\
\ding{183} Reducing delay and improving hardware efficiency 
of transcendental 
function designs by leveraging  LD bit-streams. 
\\
\ding{184} Mitigation of computational error and hardware cost in SC-based designs of transcendental 
functions. 
\\
\ding{185} Demonstration of two novel applications in SC: 
\textit{image geometric transformation} and \textit{robotic manipulator positioning}. 
\\

\begin{figure}[t]
  \centering
  \includegraphics[width=240pt]{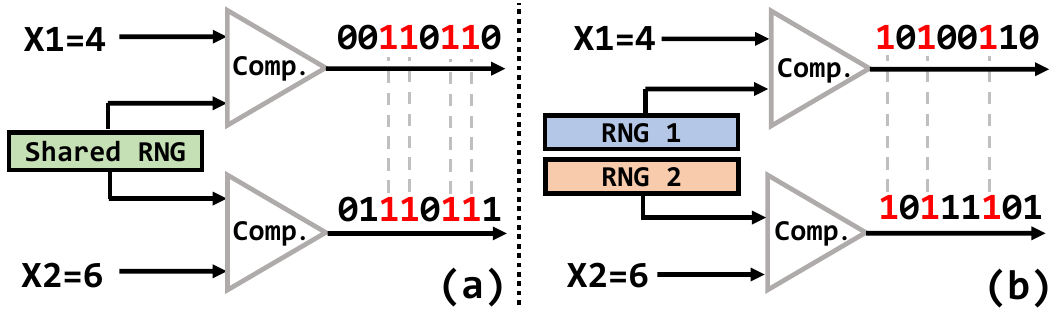}
  \vspace{-1.0em}
  \caption{Bit-stream generation using RNG and comparator (a) Shared RNG for correlated bit-streams. (b) 
  Independent RNGs for uncorrelated bit-streams.}
  \label{shared_unshared_RNG}
  \vspace{-1.5em}
\end{figure} 

\vspace{0.5em}
The remainder of this manuscript is organized as follows. Section~\ref{background} reviews the fundamentals of SC. 
Section~\ref{dse} discusses core concepts for implementing 
transcendental 
functions, transitioning from traditional SC approaches. 
Section~\ref{framework} introduces the proposed framework, featuring a novel random number 
generation technique and improvements over state-of-the-art (SOTA) designs. 
Section~\ref{experiments} presents detailed
experimental results, including accuracy and hardware cost comparisons. Section~\ref{proof_concept} provides validation and justification of the proposed approach relative to SOTA methods.
Section~\ref{design} demonstrates 
the application of SC through two representative use cases. 
{Section~\ref{discussion} provides a detailed discussion and comparison with SOTA design implementations.
}
Finally, Section~\ref{conclusion} concludes the paper by summarizing the key contributions.


\section{Fundamentals of SC}
\label{background}
SC offers an unconventional approach to computation by departing from the traditional binary radix system that encodes information positionally~\cite{10.1145/2465787.2465794}. 
Rather than relying on fixed-point or floating-point representations, which demand complex arithmetic circuits, SC realizes arithmetic with simple bitwise logic over probabilistic bit-streams. 
For example, bit-wise \texttt{AND} implements the multiplication operation when applied to independent bit-streams. 
Other fundamental SC arithmetic circuits  include a \texttt{MUX} for scaled addition ($add$)~\cite{10.1145/3491213}, an inverted-input \texttt{MUX} for scaled subtraction ($sub$), 
and a \texttt{JK-FF}-based design for approximate division ($div$)~\cite{8671762}. 

Any scalar value within the range {$0\leq X\leq N$} (or equivalently {$0 \leq \frac{X}{N} \leq 1$} when normalized to the unit interval) can be encoded into an $N$-bit stochastic bit-stream using a hardware configuration composed of an RNG and a comparator. For each bit position, the RNG generates a random value ($R$), and the comparator evaluates the condition {$X > R_i$}. If true, the corresponding bit position is set to logic-$1$; otherwise, it is set to logic-$0$. This bit-stream generation process is a fundamental step in SC systems.

The performance of SC operations 
directly depends on the BSG unit. The quality of randomness and the degree of correlation between bit-streams play a pivotal role in the performance and accuracy of computations. 
Let $X1$ and $X2$ denote scalar values encoded into stochastic bit-streams. Existing approaches to bit-stream generation predominantly adopt one of two methodologies: 
I) \textit{\textit{shared} RNG method}: a single random source is shared with two comparators to produce two \textit{correlated} bit-streams (Fig.~\ref{shared_unshared_RNG}(a)), II) \textit{\textit{independent} RNG method}: two separate random sources are utilized to generate two \textit{uncorrelated} bit-streams (Fig.~\ref{shared_unshared_RNG}(b)). 

Two primary BSG setups stand out in the state-of-the-art SC designs: 
i) 
a linear-feedback shift register (LFSR) combined with a comparator~\cite{7459560} 
and ii) 
a Sobol sequence generator combined with a comparator~\cite{Najafi_TVLSI_2019}. 
The former produces pseudo-random sequences, while the latter generates quasi-random sequences. Pseudo-random sources use deterministic mathematical algorithms to produce sequences that repeat when initialized with the same seed. 
Quasi-randomness introduces
structured variability, generating sequences that mimic randomness while maintaining a more uniform distribution.
These sequences are characterized by their evenly dispersed points in space, leading to better uniformity.
We later discuss that, for this work, we leverage quasi-random sequences by adopting cost-effective \textit{Van der Corput} (VDC) sequence generators. These provide a uniform distribution with favorable hardware properties.

The CLB unit processes the generated stochastic bit-streams. Any arithmetic function $f$ can be formulated in terms of \textit{probabilities} associated with the input bit-streams. 
For a bit-stream of length $N$, the probability of observing logic-$1$ is given by {\footnotesize $P = \frac{\sum{logic-1}}{N}$}. For two input operands $X1$ and $X2$, the corresponding probabilities are denoted by $P_{X1}$ and $P_{X2}$. 
The output of a stochastic  function $f$ is then expressed as 
$f_{mul} = P_{X1} \times P_{X2}$, $f_{add} = P_{X1} + P_{X2}$, $f_{sub} = P_{X1} - P_{X2}$, and {\footnotesize $f_{div} = \frac{P_{X1}}{P_{X2}}$}. 
In practice, discrepancies may arise between the \textit{expected} and the \textit{produced} outputs of SC systems due to three primary factors: 
\ding{192} improper cross-correlation between input bit-streams, \ding{193} random fluctuations in bit-streams stemming typically from the pseudo-random sources, 
and \ding{194} functional approximations introduced by CLB.

\textit{Correlation} 
is defined by the alignment of `1's between two bit-streams.  Ensuring proper cross-correlation between input operands is essential to achieving the desired stochastic function. 
A widely used  metric 
for quantifying  the correlation between two bit-streams is the \textit{Stochastic Cross Correlation} ($SCC$)~\cite{Alaghi_SCC1}:
\vspace{-1em}

\begin{equation}
\label{SCC_equation}
    SCC = \begin{cases} \frac{ad-bc}{N \times min(a+b, a+c)-(a+b) \times(a+c)} & ,  \  if \ ad>bc \\ \frac{ad-bc}{(a+b) \times(a+c) - N \times max(a-d, 0)} & ,  \ else\end{cases}
\end{equation}
where, $a$, $b$, $c$, and $d$ represent the logic pairs $11$, $10$, $01$, and $00$, respectively, taken from the corresponding bit positions of the two bit-streams. 

The level of correlation is determined by the number of overlapping logic-1s across all bit positions.
When  this overlap is maximized, the bit-streams exhibit 
\textit{maximum correlation} ($SCC\!\!=\!\!+1$); when the overlap is minimized, they are in \textit{minimum correlation} ($SCC\!\!=\!\!-1$).
A special case occurs when the bit-streams have \textit{zero correlation} ($SCC\!\!=\!0$), 
which is desirable in certain SC operations such as multiplication using bit-wise \texttt{AND} operation.
Let $X1 = 4$, $X2 = 6$, and $N = 8$. The input probability values are $P_{X1} = \frac{4}{8}$ and $P_{X2} = \frac{6}{8}$. The three correlation measures for bit-streams are exemplified as follows:

{
\noindent
\\
{
{\ding{172} { $\begin{smallmatrix} X1 & \rightarrow & \textcolor{black}0 & \textcolor{black}0 & \textcolor{black}0 & \textcolor{black}0 & \textcolor{red}1 & \textcolor{red}1 & \textcolor{red}1 & \textcolor{red}1 \\ X2 & \rightarrow & \textcolor{black}0 & \textcolor{black}0 & \textcolor{black}1 & \textcolor{black}1 & \textcolor{red}1 & \textcolor{red}1 & \textcolor{red}1 & \textcolor{red}1  \end{smallmatrix}$} \ding{217} \texttt{Maximum Correlation} \ \ \ \ } 
 \\
 \\
{\ding{173}  { $\begin{smallmatrix} X1 & \rightarrow & \textcolor{black}1 & \textcolor{black}1 & \textcolor{red}1 & \textcolor{red}1 & \textcolor{black}0 & \textcolor{black}0 & \textcolor{black}0 & \textcolor{black}0 \\ X2 & \rightarrow & \textcolor{black}0 & \textcolor{black}0 & \textcolor{red}1 & \textcolor{red}1 & \textcolor{black}1 & \textcolor{black}1 & \textcolor{black}1 & \textcolor{black}1  \end{smallmatrix}$} \ding{217} \texttt{Minimum Correlation}} 
\\
\\
{\ding{174} { $\begin{smallmatrix} X1 & \rightarrow & \textcolor{black}0 & \textcolor{black}1 & \textcolor{black}0 & \textcolor{red}1 & \textcolor{black}0 & \textcolor{red}1 & \textcolor{black}0 & \textcolor{red}1 \\ X2 & \rightarrow & \textcolor{black}0 & \textcolor{black}0 & \textcolor{black}1 & \textcolor{red}1 & \textcolor{black}1 & \textcolor{red}1 & \textcolor{black}1 & \textcolor{red}1  \end{smallmatrix}$} \ding{217} \texttt{Zero Correlation}}
\\
}

The parts highlighted in red 
depict the overlapping 1 bits. 
For \texttt{Maximum Correlation} \ding{172}, the positions of overlapping `1's are aligned to the right for clarity, although they may occur randomly. For \texttt{Minimum Correlation} \ding{173}, the overlapping `1's are minimum, causing 1s in the first and the second operand to be apart. Cross-correlation impacts 
the behavior of the CLB unit, enabling different functional outcomes~\cite{9319166,9126828}. 
For example, with \texttt{Zero Correlation} \ding{174}, 
bit-wise \texttt{AND} operation between $X1$ and $X2$ yields the output {$\begin{smallmatrix} Y & \rightarrow & \textcolor{black}0 & \textcolor{black}0 & \textcolor{black}0 & \textcolor{red}1 & \textcolor{black}0 & \textcolor{red}1 & \textcolor{black}0 & \textcolor{red}1 \end{smallmatrix}$}. The expected $f$ for zero-correlated input bit-streams supplied to the \texttt{AND} gate is {\footnotesize $f_{mul} = P_{X1} \times P_{X2}$}. The probability of output $Y$ is $\frac{3}{8}$, validating the expected operation (i.e., multiplication): 
{\footnotesize $f_{mul} = P_{X1} \times P_{X2} = \frac{4}{8} \times \frac{6}{8} = \frac{3}{8}$}. 
However, the cases of \texttt{Maximum} and \texttt{Minimum Correlation} in \ding{172} and \ding{173} 
yield different output probability values after bit-wise \texttt{AND}: $Y \textsubscript{\ding{172}} = \frac{4}{8}$ and $Y \textsubscript{\ding{173}} = \frac{2}{8}$. These varying correlations for bitwise \texttt{AND} elucidate different functions, such as {\footnotesize $f \textsubscript{\ding{172}} = \min(P_{X1}, P_{X2})$} and {\footnotesize $f \textsubscript{\ding{173}}  =  \max(0, P_{X1} + P_{X2}  -  1)$}. As demonstrated, correlation plays a crucial role in defining CLB behavior, underscoring  the importance  of the BSG unit. 
In this work, we significantly enhance the performance and hardware efficiency of  SC-based transcendental function designs by revisiting the RNGs used in the BSG units.



\section{From Conventional to SC-based Transcendental Functions}
\label{dse}
\label{methodology_trigonometriSC}
A key advantage of SC is its ability to perform complex arithmetic functions with simple and low-cost 
designs. This study proposes an 
efficient approach for implementing transcendental 
functions, with a focus on improving accuracy while reducing hardware cost compared to SOTA SC designs. Previous research has explored various methods for SC-based arithmetic, including piecewise-linear approximation~\cite{8825149}, FSM-based designs~\cite{8252535}, and polynomial factorization techniques~\cite{10.1145/2902961.2902999}. More recent studies have 
emphasized the implementation of polynomial functions within CLB units, which are particularly advantageous  
for machine learning applications~\cite{9843901,9444648}. 
This work highlights the critical role of lightweight, high-quality BSG units, 
which indirectly affect the performance of CLBs. 

Transcendental functions play a critical role in diverse applications ranging from communication systems to computer vision~\cite{chen2023object, Wang2021, Ankalaki2023}. SC becomes particularly important when robustness and hardware efficiency are the primary design objectives in implementing these functions. 
Among SOTA work, Parhi and Liu~\cite{Trig-Parhi} explored SC-based implementations of complex non-linear functions, including trigonometric, exponential, logarithm, hyperbolic tangent, and sigmoid, by leveraging the truncated Maclaurin series expansions. 
Alternative SC implementations of these functions have also been proposed using Chebyshev polynomials~\cite{Kind_Approx_SC_Chebyshev_2025}, FSM~\cite{8357309}, and Bernstein Polynomials~\cite{Weikang_2011,9531838}. 
Other SOTA research has extended this domain to device-level designs beyond CMOS technology. 
For example, Chen et al.~\cite{Chen_TETC2023} employed adiabatic quantum-flux parametron (AQFP) superconducting technology in conjunction with Bernstein Polynomial structures to realize transcendental  functions. 
To the best of our knowledge, our work is the first to incorporate quasi-randomness in the BSG units of SC designs, enabling more efficient SC-based implementations of transcendental functions. 

\begin{table}[t]
\centering
\renewcommand{\arraystretch}{2}
\setlength{\tabcolsep}{4.pt}
\caption{Truncated Maclaurin Series Expansions of Transcendental Functions Explored in This Study}
\vspace{-1em}
\begin{tabular}{|c|c|} 
\hline
\textbf{Function} & \textbf{Polynomial Equivalent} \\ 
\hline
$\sin(x)$ & $x  -  \frac{x^{3}}{3!}  +  \frac{x^{5}}{5!}  -  \frac{x^{7}}{7!}$ \\ 
$\cos(x)$ & $1  -  \frac{x^{2}}{2!}  +  \frac{x^{4}}{4!}  -  \frac{x^{6}}{6!} + \frac{x^8}{8!}$ \\ 
$\tan(x)=\frac{\sin(x)}{\cos(x)}$ & $x+\frac{x^{3}}{3}+\frac{2x^{5}}{15}+\frac{17x^{7}}{315}$ \\ 
$\tanh(x)$ & $x  -  \frac{x^{3}}{3}  +  \frac{2x^{5}}{15}  -  \frac{17x^{7}}{315}$ \\ 
$\arctan(x)$ & $x  -  \frac{x^{3}}{3}  +  \frac{x^{5}}{5}  -  \frac{x^{7}}{7}$ \\ 
$sigmoid(x)$ & $\frac{1}{2}  +  \frac{x}{4}  -  \frac{x^{3}}{48}  +  \frac{x^{5}}{480}$ \\ 
$Sinc(x)=\frac{\sin(x)}{x}$ & $1  -  \frac{x^{2}}{3!}  +  \frac{x^{4}}{5!}  -  \frac{x^{6}}{7!}$ \\ 
$e^{-x}$ & $1  -  x  +  \frac{x^{ 2 }}{2!}  -  \frac{x^{ 3 }}{3!}  +  \frac{x^{ 4 }}{4!}  -  \frac{x^{ 5 }}{5!}$ \\ 
$\ln(1+x)$ & $x  -  \frac{x^{ 2 }}{2}  +  \frac{x^{ 3 }}{3}  -  \frac{x^{ 4 }}{4}  +  \frac{x^{ 5 }}{5}$ \\
\hline
\end{tabular}
\label{func_ML}
\vspace{-1em}
\end{table}
\color{black}

In mathematics, transcendental functions--unlike algebraic functions--form a distinct class that cannot be expressed as a finite combination of algebraic operations such as addition, subtraction, multiplication, division, exponentiation, or root extraction. Common examples include \textit{trigonometric} (e.g., $\sin(x)$), \textit{exponential} (e.g., $a^x$), \textit{logarithmic} (e.g., $\log_a{x}$), and \textit{hyperbolic} (e.g., $\tanh(x)$) functions. By contrast, expressions such as $x^3+1$, $\frac{2x-3}{x+1}$, and $\sqrt[5]{x}$ are classified  as algebraic functions.
Our approach significantly enhances the SC design of transcendental 
arithmetic functions, distinguishing it from existing SOTA solutions. 
We focus on hardware-efficient CMOS-based implementations for key functions such as $\mathbf{sin(x)}$, $\mathbf{cos(x)}$, $\mathbf{tan(x)}$, $\mathbf{e\textsuperscript{\textbf{$-x$}}}$, $\mathbf{tanh(x)}$, $\mathbf{arctan(x)}$, $\mathbf{sigmoid(x)}$, $\mathbf{Sinc(x)}$, and $\mathbf{ln(1+x)}$, leveraging the truncated Maclaurin series expansions and Horner's rule~\cite{Trig-Parhi,9444648}. 
The Maclaurin series of a function $f(x)$ is a special case of the Taylor series 
centered at zero, expressed as { $\sum_{n=0}^{\infty}\frac{f^{(n)}}{n!}x^{n}$,} where $f^{(n)}$ denotes the $n$-th derivative of $f(x)$. In what follows, we present approximate SC designs for the functions above using Horner's rule. 
Table~\ref{func_ML} lists the truncated Maclaurin series expansion for each function explored in this study.

\begin{figure}[t]
  \centering
  \includegraphics[width=\linewidth]{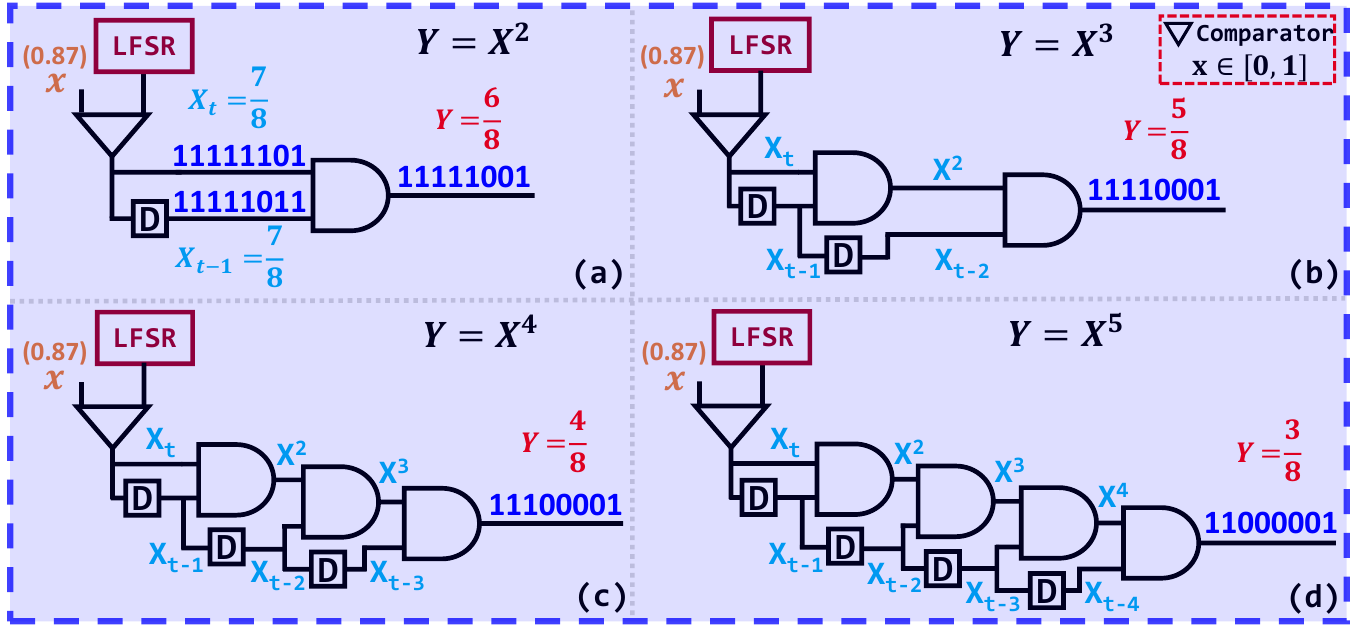}
  \vspace{-1.8em}
  \caption{Conventional designs of simple polynomial functions in SC, (a) Quadratic function ($x^2$), (b) Cubic function ($x^3$), (c) Quartic function ($x^4$), and (d) Quintic function ($x^5$).}
  \label{polynomials_fig}
  \vspace{-1.5em}
\end{figure}

\color{black}
  \vspace{-.5em}
\subsection{Polynomial Design in SC}
\label{polynomials}
To facilitate the design of complex transcendental 
functions, we start by implementing 
basic polynomial functions, including quadratic ($x^2$), cubic ($x^3$), quartic ($x^4$), quintic ($x^5$), and higher-order polynomials in SC. 
The quadratic function ($x^2$) can be implemented in SC by using a single \texttt{AND} gate combined with a decorrelator element (e.g., a one-bit shift register or a D-type flip-flop (\texttt{D-FF})). 
Similarly, the $x^3$, $x^4$, and $x^5$ functions can be implemented using two, three, and four \texttt{AND} gates paired with two, three, and four delay elements, respectively, to ensure the necessary level of independence required at the input of each \texttt{AND} gate (see Section~\ref{background}). 
Figs.~\ref{polynomials_fig}(a)-(d) illustrate the circuit designs for these polynomial functions for the bit-stream length of $N=8$.

As an example, to compute the polynomial values of $0.875$ with a bit-stream length of 8, the corresponding bit-stream $X_t=11111101$ is generated along with its one-bit shifted version $X_{t-1}=11111011$. The resulting bit-stream ($Y$) represents the expected outcome of $(0.87)^2$ as depicted in Fig.~\ref{polynomials_fig}(a). Higher-order polynomials are implemented in a similar manner by incrementally adding delay elements into the intermediate circuitry, as shown in Figs.~\ref{polynomials_fig}(b)-(d).
Building on these foundational SC implementations of simple polynomials, we explore the design 
of more complex transcendental 
functions in the following section. 
The polynomial order for each function is chosen to align with those used in conventional approaches to provide a fair comparison with our method.

\section{Proposed Framework: \Design}
\label{framework}

Our proposed framework, \Design, enhances 
the BSG unit, enabling modifications to CLBs for more efficient SC implementation of transcendental 
functions. While recent SOTA works have increasingly emphasized the design of configurable CLBs to support these 
functions--primarily targeting the function blocks~\cite{9843901,9444648,Trig-Parhi}--the role of the BSG unit has often been overlooked. 
Most existing SC designs assume BSG units rely on pseudo-random sequences generated using LFSRs (Fig.~\ref{BSG}(a)). 
However, achieving high accuracy with short bit-streams--essential for low latency and energy-efficient processing--requires innovative bit-stream generation techniques. LD sequence generators, such as Sobol 
generators~\cite{8327916, 10041889, 9781616}, offer promising alternatives by producing fast-converging quasi-random bit-streams, albeit with higher 
hardware cost (Fig.~\ref{BSG}(b)). The proposed \Design~framework adopts an alternative LD sequence 
to optimize the RNG block, thereby improving both design efficiency and accuracy. 
We demonstrate that properly employing novel LD sequences can significantly enhance hardware efficiency and computational accuracy.
In the next section, we 
introduce 
\textit{Van der Corput (VDC)} sequences as a specific type of LD quasi-random sequences, along with 
a hardware-friendly implementation. 


\begin{figure}[t]
  \centering
  \includegraphics[width=\linewidth]{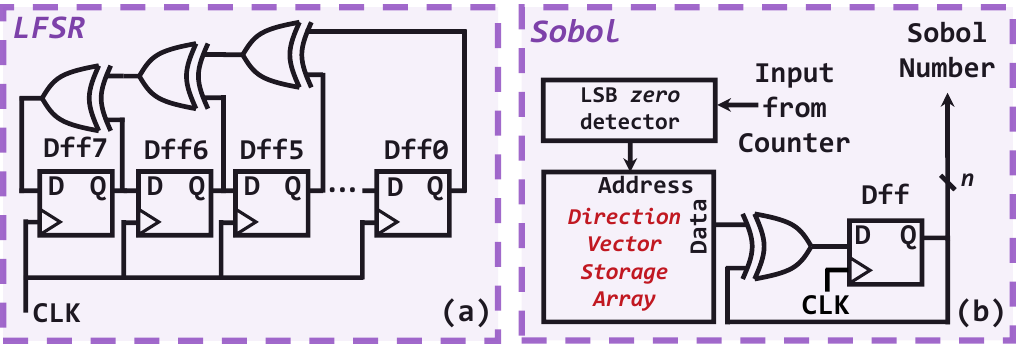}
  \vspace{-2em}
  \caption{The overall design of the state-of-the-art RNGs 
  (a) LFSR design (8-bit); \textit{maximal length LFSR with polynomial $x^8+x^7+x^6+1$ is demonstrated as an example}, and (b) Sobol design.}
  \label{BSG}
  \vspace{-0.8em}
\end{figure}

\vspace{-.5em}

\subsection{Van der Corput (VDC) Sequences}
\label{vdc}
VDC sequences are a class of random sequences widely studied in number 
and discrepancy theory, particularly recognized for their uniform distribution properties~\cite{VDC}. 
They are generated using 
the radical inverse function, defined for any base  
$\mathcal{B}\geq 2$~\cite{perm_vdc}. A defining feature of VDC sequences is their LD property, which ensures even coverage of the unit interval.  
This makes them particularly valuable in numerical integration, quasi-Monte Carlo methods, and applications in computer graphics and computational mathematics~\cite{vdc_air}. To construct a VDC sequence,  
natural numbers are mapped to the unit interval [0,1) through base $\mathcal{B}$ expansions (denoted as VDC-$\mathcal{B}$). Each term is obtained by reversing the digits of a natural number expressed in 
base $\mathcal{B}$, then interpreting the reversed digits as a fractional value. This process guarantees 
the equidistribution property modulo one. For example, the decimal number 77 is represented as 
$(302)_5$ in base-5. 
Reversing the digits yields  $(0.203)_5 = 2\times 5^{-1}+0\times 5^{-2}+3\times 5^{-3}=\frac{53}{125}=(0.424)_{10}$. 

\subsection{VDC-$2^n$: An Accurate and Hardware-Friendly Case} 
\label{vdc-2}

Inspired by the general concept of VDC sequences, we employ a hardware-friendly variant 
with  powers-of-2 radices 
(VDC-$2^n$) for the BSG unit.
Generating VDC-$2^n$ sequences 
requires only a simple modulo-$m$ up counter, where $m\geq n$. This up counter may operate in either asynchronous or synchronous mode. 
The binary output of this counter is partitioned into subgroups of $n$ bits; if the final subgroup contains fewer than $n$ bits, zero-padding is applied to ensure a complete $n$-bit. 
These subgroups are then reordered through hardwiring, such that the least significant subgroup becomes 
the most significant, and vice versa. The resulting $m$-bit binary number is subsequently mapped to the $[0,1)$ interval, producing the desired VDC-$2^n$ sequence.

Integrating VDC-$2^n$ into \Design~enables 
low-cost bit-stream generation 
using a simple $\log_{2}{N}$-bit up counter, where $N$ denotes the bit-stream length. 
The counter output can be flexibly configured for any desired powers-of-2 radix. A key advantage of VDC-$2^n$ 
is its flexibility: 
up to $\log_2N$ distinct and independent sequences can be generated simultaneously without additional hardware. 
Fig.~\ref{BSG_VDC} (a)-(d) illustrates the hardware design of VDC-2, VDC-4, VDC-8, and VDC-16, respectively. All designs share the same counter but differ in their hardwired reordering logic. 
By leveraging LD sequences 
in the form of VDC-$2^n$, \Design~achieves a lightweight and efficient BSG design, offering a compelling alternative to existing RNG designs for both the primary inputs and
scalar coefficients in the mid-stages of transcendental functions implementation. 

\begin{figure}[t]
  \centering
  \includegraphics[width=\linewidth]{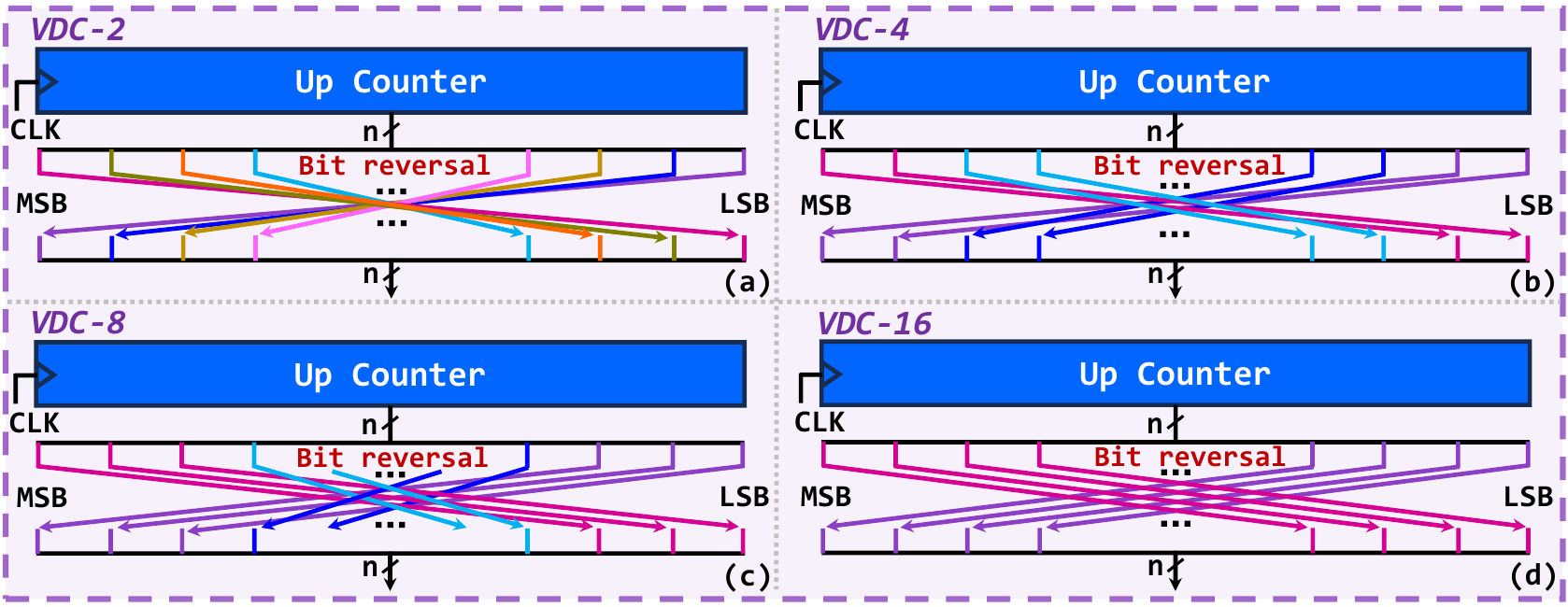}
  \vspace{-1.5em}
  \color{black}
  \caption{The overall designs of VDC-$2^n$ RNGs using 
  binary up counter (a) VDC-$2$; reversing every single bit of the counter output, 
  (b) VDC-$4$; reversing every 2-bit subgroup of counter outputs, (c) VDC-$8$; reversing every 3-bit subgroup of counter outputs, and (d) VDC-$16$; reversing every 4-bit subgroup of counter outputs. All designs are considered as $n$-bit precision - the same circuit is used with different hardwiring.}
  \color{black}
  \label{BSG_VDC}
  \vspace{-1.em}
\end{figure}

\begin{figure}[t]
  \centering
  \includegraphics[width=\linewidth]{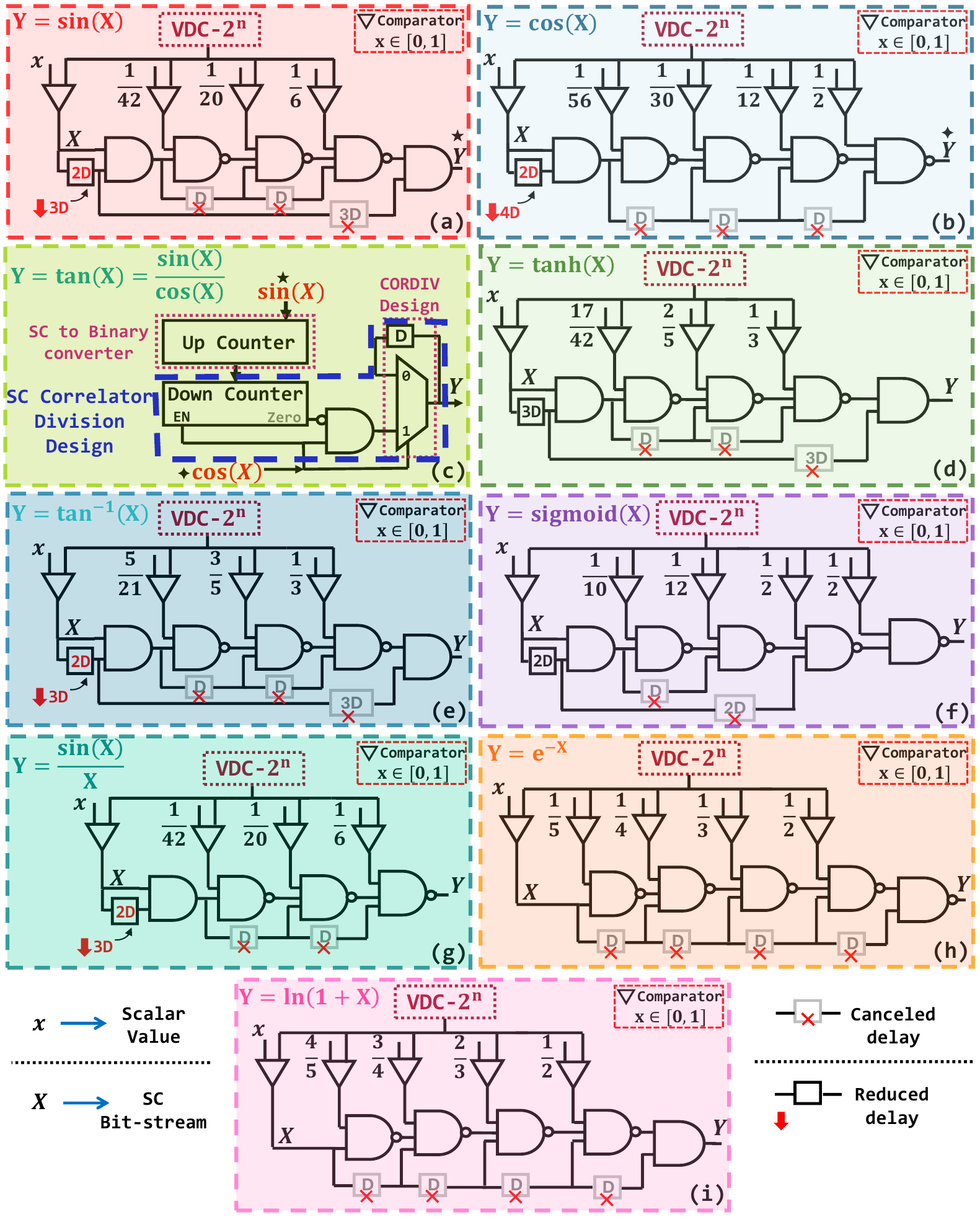}
  \vspace{-1.5em}
  \caption{ Proposed \Design~framework for transcendental functions: Eliminating the delay elements by utilizing efficient BSG units. (a) $\mathbf{sin( x )}$, (b) $\mathbf{cos( x )}$, (c) $\mathbf{tan( x )}$, 
  (d) $\mathbf{tanh( x )}$, (e) $\mathbf{arctan( x )}$, (f) $\mathbf{sigmoid( x )}$, (g) $\mathbf{Sinc( x )}$, (h) $\mathbf{e^{ - x }}$, and (i) $\mathbf{ln( 1  +  x )}$.
  }
  \label{designs}
  \vspace{-1.5em}
\end{figure}

\begin{table}[!t]
\centering
\caption{Accuracy Evaluation of \Design~and SOTA Methods for Implementing Transcendental Functions-Part1 \\(N: Bit-stream length)
}
\vspace{-1em}

\setlength\dashlinedash{0.2pt}
 \setlength\dashlinegap{1.5pt}
 \setlength\arrayrulewidth{0.3pt}
 \setlength{\tabcolsep}{0.6pt} 
 \relscale{0.74}
\begin{tabular}{|c|c|c|c|c|c|c|c|c|c|} 
\hline
\multirow{3}{*}{\begin{tabular}[c]{@{}c@{}} \textbf{Function}\end{tabular}} & \multirow{3}{*}{\begin{tabular}[c]{@{}c@{}}\textbf{Design}\\ \textbf{Approach}\end{tabular}} & \multirow{3}{*}{\begin{tabular}[c]{@{}c@{}} \textbf{N} \end{tabular}} & \multicolumn{2}{c|}{\textbf{RNG Sources}} & \multicolumn{4}{c|}{\begin{tabular}[c]{@{}c@{}} \textbf{No. of}\\ \textbf{Delay Elements} \end{tabular}} & \multirow{3}{*}{\begin{tabular}[c]{@{}c@{}} \textbf{MSE}\\ (\texttimes 10\textsuperscript{-4})\end{tabular}} \\ 
\cline{4-9}
  &  &  & \begin{tabular}[c]{@{}c@{}}\textbf{Input} 
 \end{tabular} & \begin{tabular}[c]{@{}c@{}}\textbf{Coefficients} 
 \end{tabular} & \begin{tabular}[c]{@{}c@{}}\textbf{1\textsuperscript{st}}\\ \textbf{stage}\end{tabular} & \begin{tabular}[c]{@{}c@{}}\textbf{2\textsuperscript{nd}}\\ \textbf{stage}\end{tabular} & \begin{tabular}[c]{@{}c@{}}\textbf{3\textsuperscript{rd}}\\ \textbf{stage}\end{tabular} & \begin{tabular}[c]{@{}c@{}}\textbf{4\textsuperscript{th}}\\ \textbf{stage}\end{tabular} &  \\ 
\hline 
\multirow{10}{*}{\begin{tabular}[c]{@{}c@{}} \rotatebox[origin=l]{0}{$\sin(x)$} \end{tabular}} & \begin{tabular}[c]{@{}c@{}} \Design
\end{tabular}\textsuperscript{\ding{93}} & \begin{tabular}[c]{@{}c@{}} 1024 \\ 512 \\ 256\\ 128\\ 64 \end{tabular} & \begin{tabular}[c]{@{}c@{}}VDC-4 \end{tabular} & \begin{tabular}[c]{@{}c@{}}VDC-128,256,512\\VDC-128,256,512\\VDC-128\\ VDC-2,8,128\\ VDC-2,8,8  \end{tabular} & \textbf{2} & \textbf{0} & \textbf{0} & \textbf{0} & \begin{tabular}[c]{@{}c@{}} \textbf{0.523} \\ 0.582 \\0.576\\ 0.967\\ 1.743
 \end{tabular} \\ 
\cdashline{2-10}[1pt/1pt]

 & \begin{tabular}[c]{@{}c@{}} \Design
\textsuperscript{\ding{70}} \end{tabular} &\begin{tabular}[c]{@{}c@{}} 1024\\ 512\\ 256\\ 128\\ 64 \end{tabular}& \begin{tabular}[c]{@{}c@{}}VDC-2 \end{tabular} & \begin{tabular}[c]{@{}c@{}}VDC-1024 \\ VDC-512 \\ VDC-256\\ VDC-128\\ VDC-64 \end{tabular} & 1 & \textbf{0} & \textbf{0} & \textbf{0} & \begin{tabular}[c]{@{}c@{}} 0.888\\ 0.999\\ 1.221\\ 1.894\\ 5.373 \end{tabular} \\ 
\cdashline{2-10}[1pt/1pt]
& \begin{tabular}[c]{@{}c@{}} \textbf{Work in} \textbf{\cite{Trig-Parhi}} 
\end{tabular} & \begin{tabular}[c]{@{}c@{}} 1024 \end{tabular} & \begin{tabular}[c]{@{}c@{}}Sobol-1 \end{tabular} & \begin{tabular}[c]{@{}c@{}}Sobol-2  \end{tabular} & 3 & 1 & 1 & 3 & \begin{tabular}[c]{@{}c@{}} 3.133
 \end{tabular} \\ 
\cdashline{2-10}[1pt/1pt]
& \begin{tabular}[c]{@{}c@{}} \textbf{Work in} \textbf{\cite{9444648}}
\end{tabular} & \begin{tabular}[c]{@{}c@{}} 1024 \end{tabular} & \begin{tabular}[c]{@{}c@{}}Sobol-1 \end{tabular} & \begin{tabular}[c]{@{}c@{}}Sobol-2  \end{tabular} & 1 & 1 & 1 & 1 & \begin{tabular}[c]{@{}c@{}} 0.917
 \end{tabular} \\
\cdashline{2-10}[1pt/1pt]
 &\begin{tabular}[c]{@{}c@{}} \textbf{Work in} \textbf{\cite{Trig-Parhi}} \end{tabular}  & 1024 & LFSR1 & LFSR2 & 3 & 1 & 1 & 3 & 2.256 \\ 
\cdashline{2-10}[1pt/1pt]
 &\begin{tabular}[c]{@{}c@{}} \textbf{Work in} \textbf{\cite{9444648}} \end{tabular} & 1024 & LFSR1 & LFSR2 & 1 & 1 & 1 & 1 & 5.036 \\ 
\hline
\multirow{10}{*}{\begin{tabular}[c]{@{}c@{}} \rotatebox[origin=l]{0}{$\cos(x)$} \end{tabular}} & \begin{tabular}[c]{@{}c@{}}\Design\textsuperscript{\ding{93}} \end{tabular}  &\begin{tabular}[c]{@{}c@{}} 1024 \\ 512 \\ 256\\ 128\\ 64 \end{tabular}& \begin{tabular}[c]{@{}c@{}} VDC-8 \end{tabular} & \begin{tabular}[c]{@{}c@{}} VDC-8,4,16,256\\ VDC-8,4,16,256\\ VDC-8,4,16,256\\ VDC-8,4,16,128\\ VDC-8,4,16,64 \end{tabular} & \textbf{2} & \textbf{0} & \textbf{0} & \textbf{0} & \begin{tabular}[c]{@{}c@{}} \textbf{1.073} \\ 1.087 \\ 1.096\\ 1.119\\ 3.815 \end{tabular} \\ 
\cdashline{2-10}[1pt/1pt]
 & \begin{tabular}[c]{@{}c@{}}\Design\textsuperscript{\ding{70}} \end{tabular}  &\begin{tabular}[c]{@{}c@{}} 1024 \\ 512\\ 256\\ 128\\ 64 \end{tabular}& \begin{tabular}[c]{@{}c@{}} VDC-4 \end{tabular} & \begin{tabular}[c]{@{}c@{}}VDC-16,4,8,32 \end{tabular} & 1 & \textbf{0} & \textbf{0} & \textbf{0} & \begin{tabular}[c]{@{}c@{}} 1.342 \\ 1.374\\ 1.352\\ 1.678\\ 1.484 \end{tabular} \\ 
\cdashline{2-10}[1pt/1pt]
& \begin{tabular}[c]{@{}c@{}} \textbf{Work in} \textbf{\cite{Trig-Parhi}} 
\end{tabular} & \begin{tabular}[c]{@{}c@{}} 1024 \end{tabular} & \begin{tabular}[c]{@{}c@{}}Sobol-1 \end{tabular} & \begin{tabular}[c]{@{}c@{}}Sobol-2  \end{tabular} & 4 & 1 & 1 & 1 & \begin{tabular}[c]{@{}c@{}} 42.243
 \end{tabular} \\ 
\cdashline{2-10}[1pt/1pt]
& \begin{tabular}[c]{@{}c@{}} \textbf{Work in} \textbf{\cite{9444648}}
\end{tabular} & \begin{tabular}[c]{@{}c@{}} 1024 \end{tabular} & \begin{tabular}[c]{@{}c@{}}Sobol-1 \end{tabular} & \begin{tabular}[c]{@{}c@{}}Sobol-2  \end{tabular} & 1 & 1 & 1 & 1 & \begin{tabular}[c]{@{}c@{}} 1.343
 \end{tabular} \\
\cdashline{2-10}[1pt/1pt]
 &\begin{tabular}[c]{@{}c@{}} \textbf{Work in} \textbf{\cite{Trig-Parhi}} \end{tabular}  & 1024 & LFSR1 & LFSR2 & 4 & 1 & 1 & 1 & 2.518 \\ 
\cdashline{2-10}[1pt/1pt]
 &\begin{tabular}[c]{@{}c@{}} \textbf{Work in} \textbf{\cite{9444648}} \end{tabular}  & 1024 & LFSR1 & LFSR2 & 1 & 1 & 1 & 1 & 4.934 \\
\hline
\multirow{12}{*}{\begin{tabular}[c]{@{}c@{}} \rotatebox[origin=l]{0}{$\tan(x)$} \end{tabular}} & \begin{tabular}[c]{@{}c@{}}\Design\textsuperscript{\ding{93}} \end{tabular}  &\begin{tabular}[c]{@{}c@{}}1024\\ 512\\ 256\\ 128 \end{tabular}& \begin{tabular}[c]{@{}c@{}} { $\sin(x)$:VDC-8} \\ { $\cos(x)$:VDC-4} \end{tabular} & \begin{tabular}[c]{@{}c@{}} { $\sin(x)$:}\\VDC-128,128,128 \\ { $\cos(x)$:}\\VDC-16,8,2,128 \end{tabular} & \begin{tabular}[c]{@{}c@{}} 3\\ 3 \end{tabular}& \begin{tabular}[c]{@{}c@{}} \textbf{0}\\ \textbf{0} \end{tabular}& \begin{tabular}[c]{@{}c@{}} \textbf{0}\\ \textbf{0} \end{tabular}& \begin{tabular}[c]{@{}c@{}} \textbf{1}\\ 2 \end{tabular}& \begin{tabular}[c]{@{}c@{}} \textbf{0.721} \\ 0.780 \\ 1.099\\ 1.349 \end{tabular} \\ 
\cdashline{2-10}[1pt/1pt]
 & \begin{tabular}[c]{@{}c@{}}\Design\textsuperscript{\ding{70}} \end{tabular}  &\begin{tabular}[c]{@{}c@{}} 1024\\ 512\\ 256 \end{tabular}& \begin{tabular}[c]{@{}c@{}} { $\sin(x)$:VDC-32}\\ { $\cos(x)$:VDC-8} \end{tabular} & \begin{tabular}[c]{@{}c@{}} { $\sin(x)$:}\\VDC-256,256,16  \\ { $\cos(x)$:}\\VDC-256,256,256,256 \end{tabular} &\begin{tabular}[c]{@{}c@{}} 1\\ 1 \end{tabular}&\begin{tabular}[c]{@{}c@{}} 0\\ 0 \end{tabular}&\begin{tabular}[c]{@{}c@{}} 0\\ 0 \end{tabular}&\begin{tabular}[c]{@{}c@{}} 2\\ 0 \end{tabular}& \begin{tabular}[c]{@{}c@{}} 5.471\\ 3.938\\ 4.942 \end{tabular} \\ 
\cdashline{2-10}[1pt/1pt]
& \begin{tabular}[c]{@{}c@{}} \textbf{Work in} \textbf{\cite{Trig-Parhi}} 
\end{tabular} & \begin{tabular}[c]{@{}c@{}} 1024 \end{tabular} & \begin{tabular}[c]{@{}c@{}}  $\sin(x)$:Sobol-1\\ $\cos(x)$:Sobol-1  \end{tabular} & \begin{tabular}[c]{@{}c@{}}  $\sin(x)$:Sobol-2\\ $\cos(x)$:Sobol-2  \end{tabular} &\begin{tabular}[c]{@{}c@{}} 3\\ 4 \end{tabular}&\begin{tabular}[c]{@{}c@{}} 1\\ 1 \end{tabular}&\begin{tabular}[c]{@{}c@{}} 1\\1 \end{tabular}& \begin{tabular}[c]{@{}c@{}} 1\\1 \end{tabular}& \begin{tabular}[c]{@{}c@{}} 124.403
 \end{tabular} \\ 
\cdashline{2-10}[1pt/1pt]
& \begin{tabular}[c]{@{}c@{}} \textbf{Work in} \textbf{\cite{9444648}}
\end{tabular} & \begin{tabular}[c]{@{}c@{}} 1024 \end{tabular} & \begin{tabular}[c]{@{}c@{}} $\sin(x)$:Sobol-1\\ $\cos(x)$:Sobol-1 \end{tabular} & \begin{tabular}[c]{@{}c@{}} $\sin(x)$:Sobol-2\\ $\cos(x)$:Sobol-2 \end{tabular} & 1 & 1 & 1 & 1 & \begin{tabular}[c]{@{}c@{}} 89.835
 \end{tabular} \\
\cdashline{2-10}[1pt/1pt]
 &\begin{tabular}[c]{@{}c@{}} \textbf{Work in} \textbf{\cite{Trig-Parhi}} \end{tabular} & 1024 & \begin{tabular}[c]{@{}c@{}}  $\sin(x)$:LFSR1\\$\cos(x)$:LFSR1  \end{tabular}& \begin{tabular}[c]{@{}c@{}}  $\sin(x)$:LFSR1\\$\cos(x)$:LFSR2  \end{tabular}&\begin{tabular}[c]{@{}c@{}} 3\\4 \end{tabular}&\begin{tabular}[c]{@{}c@{}} 1\\1 \end{tabular}&\begin{tabular}[c]{@{}c@{}} 1\\1 \end{tabular}&\begin{tabular}[c]{@{}c@{}} 3\\1 \end{tabular}& 9.845 \\ 
\cdashline{2-10}[1pt/1pt]
 &\begin{tabular}[c]{@{}c@{}} \textbf{Work in} \textbf{\cite{9444648}} \end{tabular} & 1024 & \begin{tabular}[c]{@{}c@{}} LFSR1 \end{tabular} & \begin{tabular}[c]{@{}c@{}} LFSR2 \end{tabular} &\begin{tabular}[c]{@{}c@{}} 1 \end{tabular}&\begin{tabular}[c]{@{}c@{}} 1 \end{tabular}&\begin{tabular}[c]{@{}c@{}} 1 \end{tabular}&\begin{tabular}[c]{@{}c@{}} 1 \end{tabular}& 4.564 \\
\hline
\multirow{10}{*}{\begin{tabular}[c]{@{}c@{}} \rotatebox[origin=l]{0}{\textbf{$\tanh(x)$}} \end{tabular}} & \begin{tabular}[c]{@{}c@{}}\Design\textsuperscript{\ding{93}} \end{tabular} &\begin{tabular}[c]{@{}c@{}} 1024 \\ 512 \\ 256\\ 128\\ 64 \end{tabular}& \begin{tabular}[c]{@{}c@{}} VDC-16 \end{tabular} & VDC-32,16,2 & 3 & \textbf{0} & \textbf{0} & \textbf{0} & \begin{tabular}[c]{@{}c@{}} \textbf{2.881} \\ 2.944 \\ 3.273\\ 3.244\\ 5.800 \end{tabular} \\ 
\cdashline{2-10}[1pt/1pt]
 & \begin{tabular}[c]{@{}c@{}}\Design\textsuperscript{\ding{70}} \end{tabular}  &\begin{tabular}[c]{@{}c@{}} 1024 \\ 512\\ 256\\ 128\\ 64 \end{tabular}& \begin{tabular}[c]{@{}c@{}} VDC-2 \end{tabular} & \begin{tabular}[c]{@{}c@{}}VDC-4,64,64 \end{tabular} & 1 & \textbf{0} & \textbf{0} & \textbf{0} & \begin{tabular}[c]{@{}c@{}} 3.164 \\ 3.172\\ 3.190\\ 3.240\\ 3.392 \end{tabular} \\

\cdashline{2-10}[1pt/1pt]
 &\begin{tabular}[c]{@{}c@{}} \textbf{Work in} \textbf{\cite{Trig-Parhi}} \end{tabular}  & 1024 & Sobol-1 & Sobol-2 & 3 & 1 & 1 & 3 & 7.731 \\ 

\cdashline{2-10}[1pt/1pt]
 &\begin{tabular}[c]{@{}c@{}} \textbf{Work in} \textbf{\cite{9444648}} \end{tabular}  & 1024 & Sobol-1 & Sobol-2 & 1 & 1 & 1 & 1 & 2.390 \\
 
\cdashline{2-10}[1pt/1pt]
 &\begin{tabular}[c]{@{}c@{}} \textbf{Work in} \textbf{\cite{Trig-Parhi}} \end{tabular}  & 1024 & LFSR1 & LFSR2 & 3 & 1 & 1 & 3 & 6.579 \\ 
\cdashline{2-10}[1pt/1pt]
 &\begin{tabular}[c]{@{}c@{}} \textbf{Work in} \textbf{\cite{9444648}} \end{tabular}  & 1024 & LFSR1 & LFSR2 & 1 & 1 & 1 & 1 & 12.548 \\
\hline
\end{tabular}
\vspace{-0.5em}
\justify{\scriptsize{\ding{93}: Applying \Design~to the design of~\cite{Trig-Parhi}, \ding{70}: Applying \Design~to the design of~\cite{9444648}. 
Different VDC-$2^n$ bases come from the same hardware source via simple hardwiring. 
}}
\label{acc_results}
\vspace{-1.5em}
\end{table}

\begin{table}[!t]
\centering
\caption{Accuracy Evaluation of the \Design~and SOTA Methods for Implementing Transcendental Functions-Part2 \\(N: Bit-stream length)
}
\vspace{-1em}

\setlength\dashlinedash{0.2pt}
 \setlength\dashlinegap{1.5pt}
 \setlength\arrayrulewidth{0.3pt}
 \setlength{\tabcolsep}{0.8pt}
 \relscale{0.74}
\begin{tabular}{|c|c|c|c|c|c|c|c|c|c|} 
\hline
\multirow{3}{*}{\begin{tabular}[c]{@{}c@{}} \textbf{Function}\end{tabular}} & \multirow{3}{*}{\begin{tabular}[c]{@{}c@{}}\textbf{Design}\\ \textbf{Approach}\end{tabular}} & \multirow{3}{*}{\begin{tabular}[c]{@{}c@{}} \textbf{N} \end{tabular}} & \multicolumn{2}{c|}{\textbf{RNG Sources}} & \multicolumn{4}{c|}{\begin{tabular}[c]{@{}c@{}} \textbf{No. of}\\ \textbf{Delay Elements} \end{tabular}} & \multirow{3}{*}{ \begin{tabular}[c]{@{}c@{}} \textbf{MSE}\\ (\texttimes 10\textsuperscript{-4})\end{tabular}} \\ 
\cline{4-9}
  &  &  & \begin{tabular}[c]{@{}c@{}}\textbf{Input} 
 \end{tabular} & \begin{tabular}[c]{@{}c@{}}\textbf{Coefficients} 
 \end{tabular} & \begin{tabular}[c]{@{}c@{}}\textbf{1\textsuperscript{st}}\\ \textbf{stage}\end{tabular} & \begin{tabular}[c]{@{}c@{}}\textbf{2\textsuperscript{nd}}\\ \textbf{stage}\end{tabular} & \begin{tabular}[c]{@{}c@{}}\textbf{3\textsuperscript{rd}}\\ \textbf{stage}\end{tabular} & \begin{tabular}[c]{@{}c@{}}\textbf{4\textsuperscript{th}}\\ \textbf{stage}\end{tabular} &  \\ 
\hline

\multirow{10}{*}{\begin{tabular}[c]{@{}c@{}} \rotatebox[origin=l]{0}{$sigmoid(x)$} \end{tabular}} & \begin{tabular}[c]{@{}c@{}}\Design\textsuperscript{\ding{93}} \end{tabular}  &\begin{tabular}[c]{@{}c@{}} 1024 \\ 512 \\ 256\\ 128\\ 64 \end{tabular}& \begin{tabular}[c]{@{}c@{}} VDC-1024\\ VDC-512\\ VDC-256 \end{tabular} & VDC-2,4,32 & 2 & \textbf{0} & \textbf{0} & - & \begin{tabular}[c]{@{}c@{}} \textbf{0.072} \\ 0.099 \\ 0.418\\ 2.109\\ 9.648 \end{tabular} \\ 
\cdashline{2-10}[1pt/1pt]
 & \begin{tabular}[c]{@{}c@{}}\Design\textsuperscript{\ding{70}} \end{tabular}  &\begin{tabular}[c]{@{}c@{}} 1024 \\ 512\\ 256\\ 128\\ 64 \end{tabular}& \begin{tabular}[c]{@{}c@{}} VDC-128 \end{tabular} & \begin{tabular}[c]{@{}c@{}}VDC-4 \end{tabular} & 1 & \textbf{0} & \textbf{0} & - & \begin{tabular}[c]{@{}c@{}} 0.151 \\ 0.152\\ 0.153\\ 0.187\\ 0.547  \end{tabular} \\ 
\cdashline{2-10}[1pt/1pt]

 &\begin{tabular}[c]{@{}c@{}} \textbf{Work in} \textbf{\cite{Trig-Parhi}} \end{tabular}  & 1024 & Sobol-1 & Sobol-2 & 2 & 1 & 2 & - & 189.310 \\ 
\cdashline{2-10}[1pt/1pt]

 &\begin{tabular}[c]{@{}c@{}} \textbf{Work in} \textbf{\cite{9444648}} \end{tabular}  & 1024 & Sobol-1 & Sobol-2 & 1 & 1 & 1 & - & 282.834 \\
\cdashline{2-10}[1pt/1pt]

 &\begin{tabular}[c]{@{}c@{}} \textbf{Work in} \textbf{\cite{Trig-Parhi}} \end{tabular}  & 1024 & LFSR1 & LFSR2 & 2 & 1 & 2 & - & 2.903 \\ 
\cdashline{2-10}[1pt/1pt]
 &\begin{tabular}[c]{@{}c@{}} \textbf{Work in} \textbf{\cite{9444648}} \end{tabular}  & 1024 & LFSR1 & LFSR2 & 1 & 1 & 1 & - & 0.383 \\
\hline
\multirow{10}{*}{\begin{tabular}[c]{@{}c@{}} \rotatebox[origin=l]{0}{$e$\textsuperscript{$-x$}} \end{tabular}} & \begin{tabular}[c]{@{}c@{}}\Design\textsuperscript{\ding{93}} \end{tabular}  &\begin{tabular}[c]{@{}c@{}} 1024 \\ 512 \\ 256\\ 128\\ 64 \end{tabular}& \begin{tabular}[c]{@{}c@{}} VDC-128\\ VDC-128\\ VDC-32\\ VDC-8\\ VDC-8 \end{tabular} &\begin{tabular}[c]{@{}c@{}} VDC-16,1024,512,512\\ VDC-128,512,512,512\\ VDC-32,256,256,256\\ VDC-64,128,128,128\\ VDC-64,64,128,128 \end{tabular}& \textbf{0} & \textbf{0} & \textbf{0} & \textbf{0} & \begin{tabular}[c]{@{}c@{}} 3.032 \\ 3.517 \\ 2.508\\ 3.738\\ 2.391 \end{tabular} \\ 
\cdashline{2-10}[1pt/1pt]
 & \begin{tabular}[c]{@{}c@{}}\Design\textsuperscript{\ding{70}} \end{tabular}  &\begin{tabular}[c]{@{}c@{}} 1024 \\ 512\\ 256\\ 128\\ 64 \end{tabular}& \begin{tabular}[c]{@{}c@{}} VDC-32\\ VDC-32\\ VDC-32\\ VDC-32\\ VDC-32 \end{tabular} & \begin{tabular}[c]{@{}c@{}} VDC-64\\ VDC-64\\ VDC-64\\ VDC-4\\ VDC-64 \end{tabular} & \textbf{0} & \textbf{0} & \textbf{0} & \textbf{2} & \begin{tabular}[c]{@{}c@{}} \textbf{1.730} \\ 2.647\\ 3.412\\ 2.996\\ 2.871  \end{tabular} \\ 
\cdashline{2-10}[1pt/1pt]

 &\begin{tabular}[c]{@{}c@{}} \textbf{Work in} \textbf{\cite{Trig-Parhi}} \end{tabular}  & 1024 & Sobol-1 & Sobol-2 & 1 & 1 & 1 & 1 & 22.231 \\ 
\cdashline{2-10}[1pt/1pt]

 &\begin{tabular}[c]{@{}c@{}} \textbf{Work in} \textbf{\cite{9444648}} \end{tabular}  & 1024 & Sobol-1 & Sobol-2 & 1 & 1 & 1 & 1 & 21.210 \\
\cdashline{2-10}[1pt/1pt]

 &\begin{tabular}[c]{@{}c@{}} \textbf{Work in} \textbf{\cite{Trig-Parhi}} \end{tabular}  & 1024 & LFSR1 & LFSR2 & 1 & 1 & 1 & 1 & 9.331 \\ 
\cdashline{2-10}[1pt/1pt]
 &\begin{tabular}[c]{@{}c@{}} \textbf{Work in} \textbf{\cite{9444648}} \end{tabular}  & 1024 & LFSR1 & LFSR2 & 1 & 1 & 1 & 1 & 6.807 \\
\hline
\multirow{10}{*}{\begin{tabular}[c]{@{}c@{}} \rotatebox[origin=l]{0}{$\arctan$($x$)} \end{tabular}} & \begin{tabular}[c]{@{}c@{}}\Design\textsuperscript{\ding{93}} \end{tabular}  &\begin{tabular}[c]{@{}c@{}} 1024 \\ 512\\ 256\\ 128\\ 64 \end{tabular}& \begin{tabular}[c]{@{}c@{}} VDC-8 \end{tabular} &\begin{tabular}[c]{@{}c@{}} VDC-512,8,256\\ VDC-512,8,256\\ VDC-16,8,64\\ VDC-128,8,64\\ VDC-16,8,64 \end{tabular}& \textbf{2} & \textbf{0} & \textbf{0} & \textbf{0} & \begin{tabular}[c]{@{}c@{}} \textbf{0.835} \\ 0.854\\ 2.544\\ 2.600\\ 3.271 \end{tabular} \\ 
\cdashline{2-10}[1pt/1pt]
 & \begin{tabular}[c]{@{}c@{}}\Design\textsuperscript{\ding{70}} \end{tabular}  &\begin{tabular}[c]{@{}c@{}} 1024 \\ 512\\ 256\\ 128\\ 64 \end{tabular}& \begin{tabular}[c]{@{}c@{}} VDC-4 \end{tabular} & \begin{tabular}[c]{@{}c@{}} VDC-4,2,2 \end{tabular} & 1 & \textbf{0} & \textbf{0} & \textbf{0} & \begin{tabular}[c]{@{}c@{}} 1.733 \\ 1.853\\ 1.678\\ 2.336\\ 2.006 \end{tabular} \\ 
\cdashline{2-10}[1pt/1pt]

 &\begin{tabular}[c]{@{}c@{}} \textbf{Work in} \textbf{\cite{Trig-Parhi}} \end{tabular}  & 1024 & Sobol-1 & Sobol-2 & 3 & 1 & 1 & 3 & 8.028 \\ 
\cdashline{2-10}[1pt/1pt]

 &\begin{tabular}[c]{@{}c@{}} \textbf{Work in} \textbf{\cite{9444648}} \end{tabular}  & 1024 & Sobol-1 & Sobol-2 & 1 & 1 & 1 & 1 & 2.327 \\
\cdashline{2-10}[1pt/1pt]

 &\begin{tabular}[c]{@{}c@{}} \textbf{Work in} \textbf{\cite{Trig-Parhi}} \end{tabular}  & 1024 & LFSR1 & LFSR2 & 3 & 1 & 1 & 3 & 1.947 \\ 
\cdashline{2-10}[1pt/1pt]
 &\begin{tabular}[c]{@{}c@{}} \textbf{Work in} \textbf{\cite{9444648}} \end{tabular}  & 1024 & LFSR1 & LFSR2 & 1 & 1 & 1 & 1 & 7.518 \\
\hline
\multirow{10}{*}{\begin{tabular}[c]{@{}c@{}} \rotatebox[origin=l]{0}{$Sinc(x)$} \end{tabular}} & \begin{tabular}[c]{@{}c@{}}\Design\textsuperscript{\ding{93}} \end{tabular}  &\begin{tabular}[c]{@{}c@{}} 1024 \\ 512\\ 256\\ 128\\ 64 \end{tabular}& \begin{tabular}[c]{@{}c@{}} VDC-8 \end{tabular} &\begin{tabular}[c]{@{}c@{}} VDC-256,32,1024\\ VDC-256,32,512\\ VDC-256,32,256\\ VDC-4,64,128\\ VDC-4,8,64 \end{tabular}& \textbf{2} & \textbf{0} & \textbf{0} & - & \begin{tabular}[c]{@{}c@{}} \textbf{0.124} \\ 0.186 \\ 0.297\\ 0.386\\ 1.145 \end{tabular} \\ 
\cdashline{2-10}[1pt/1pt]
 & \begin{tabular}[c]{@{}c@{}}\Design\textsuperscript{\ding{70}} \end{tabular}  &\begin{tabular}[c]{@{}c@{}} 1024 \\ 512 \\ 256\\ 128\\ 64 \end{tabular}& \begin{tabular}[c]{@{}c@{}} VDC-4 \end{tabular} & \begin{tabular}[c]{@{}c@{}} VDC-512\\ VDC-256\\ VDC-128\\ VDC-128\\ VDC-64 \end{tabular} & 1 & \textbf{0} & \textbf{0} & - & \begin{tabular}[c]{@{}c@{}} 0.181 \\ 0.258\\ 0.365\\ 0.416\\ 2.917 \end{tabular} \\ 
\cdashline{2-10}[1pt/1pt]

 &\begin{tabular}[c]{@{}c@{}} \textbf{Work in} \textbf{\cite{Trig-Parhi}} \end{tabular}  & 1024 & Sobol-1 & Sobol-2 & 3 & 1 & 1 & - & 4.324 \\ 
\cdashline{2-10}[1pt/1pt]

 &\begin{tabular}[c]{@{}c@{}} \textbf{Work in} \textbf{\cite{9444648}} \end{tabular}  & 1024 & Sobol-1 & Sobol-2 & 1 & 1 & 1 & - & 3.450 \\
\cdashline{2-10}[1pt/1pt]

 &\begin{tabular}[c]{@{}c@{}} \textbf{Work in} \textbf{\cite{Trig-Parhi}} \end{tabular}  & 1024 & LFSR1 & LFSR2 & 3 & 1 & 1 & - & 7.081 \\ 
\cdashline{2-10}[1pt/1pt]
 &\begin{tabular}[c]{@{}c@{}} \textbf{Work in} \textbf{\cite{9444648}} \end{tabular}  & 1024 & LFSR1 & LFSR2 & 1 & 1 & 1 & - & 1.952 \\
\hline
\multirow{10}{*}{\begin{tabular}[c]{@{}c@{}} \rotatebox[origin=l]{0}{\textbf{$\ln(1+x)$}} \end{tabular}} & \begin{tabular}[c]{@{}c@{}}\Design\textsuperscript{\ding{93}} \end{tabular}  &\begin{tabular}[c]{@{}c@{}} 1024 \\ 512\\ 256\\ 128\\ 64 \end{tabular}& \begin{tabular}[c]{@{}c@{}} VDC-64\\ VDC-64\\ VDC-16\\ VDC-8\\ VDC-4 \end{tabular} &\begin{tabular}[c]{@{}c@{}} VDC-4,512,1024,512\\ VDC-4,256,512,256\\ VDC-64,128,2\\ VDC-32,64,2\\ VDC-8,32,2 \end{tabular}& \textbf{0} & \textbf{0} & \textbf{0} & \textbf{0} & \begin{tabular}[c]{@{}c@{}} 0.996 \\ 2.482\\ 1.367\\ 3.470\\ 4.973 \end{tabular} \\ 
\cdashline{2-10}[1pt/1pt]
 & \begin{tabular}[c]{@{}c@{}}\Design\textsuperscript{\ding{70}} \end{tabular}  &\begin{tabular}[c]{@{}c@{}} 1024 \\ 512\\ 256\\ 128\\ 64 \end{tabular}& \begin{tabular}[c]{@{}c@{}} VDC-16\\ VDC-16\\ VDC-16\\ VDC-16\\ VDC-4 \end{tabular} & \begin{tabular}[c]{@{}c@{}} VDC-32,64,256,256\\ VDC-32,64,256,256\\ VDC-32,64,256,256\\ VDC-32,8,128,128\\ VDC-2,32,64,64  \end{tabular} & \textbf{0} & \textbf{0} & \textbf{0} & \textbf{0} & \begin{tabular}[c]{@{}c@{}} \textbf{0.734} \\ 1.023\\ 0.979\\ 2.277\\ 1.593 \end{tabular} \\ 
\cdashline{2-10}[1pt/1pt]

 &\begin{tabular}[c]{@{}c@{}} \textbf{Work in} \textbf{\cite{Trig-Parhi}} \end{tabular}  & 1024 & Sobol-1 & Sobol-2 & 1 & 1 & 1 & 1 & 15.262 \\ 
\cdashline{2-10}[1pt/1pt]

 &\begin{tabular}[c]{@{}c@{}} \textbf{Work in} \textbf{\cite{9444648}} \end{tabular}  & 1024 & Sobol-1 & Sobol-2 & 1 & 1 & 1 & 1 & 17.258 \\
\cdashline{2-10}[1pt/1pt]

 &\begin{tabular}[c]{@{}c@{}} \textbf{Work in} \textbf{\cite{Trig-Parhi}} \end{tabular}  & 1024 & LFSR1 & LFSR2 & 1 & 1 & 1 & 1 & 4.131 \\ 
\cdashline{2-10}[1pt/1pt]
 &\begin{tabular}[c]{@{}c@{}} \textbf{Work in} \textbf{\cite{9444648}} \end{tabular}  & 1024 & LFSR1 & LFSR2 & 1 & 1 & 1 & 1 & 3.294 \\
\hline
\end{tabular}
\vspace{-0.5em}
\justify{\scriptsize{\ding{93}: Applying \Design~to the design of~\cite{Trig-Parhi}, \ding{70}: Applying \Design~to the design of~\cite{9444648}. 
Different VDC-$2^n$ bases come from the same hardware source via simple hardwiring. 
}}
\vspace{-1.7em}
\label{acc_results2}
\end{table}



\begin{table*}
\centering
\caption{Hardware Cost Comparison of \Design~and SOTA Designs for Implementing Transcendental Functions}
\vspace{-1em}
\setlength{\tabcolsep}{3.2pt}
\begin{tabular}{|c|c|c|c|c|c|c|c|c|c||c|c|c|c|c|c|c|c|} 
\hline
\multirow{2}{*}{\begin{tabular}[c]{@{}c@{}}\\ \textbf{Functions} \end{tabular}} & \multirow{2}{*}{\begin{tabular}[c]{@{}c@{}}\\ \textbf{N} \end{tabular}} & \multicolumn{4}{c|}{\textbf{\Design}\textsuperscript{\ding{93}}} & \multicolumn{4}{c||}{\textbf{\Design}\textsuperscript{\ding{70}}} & \multicolumn{4}{c|}{\textbf{Work in~\cite{Trig-Parhi}}} & \multicolumn{4}{c|}{\textbf{Work in~\cite{9444648}}} \\ 
\cline{3-18}
 &  & \begin{tabular}[c]{@{}c@{}}\textbf{Area} \\\textbf{($\mu$\textbf{m\textsuperscript{2}})}\end{tabular} & \begin{tabular}[c]{@{}c@{}}\textbf{CPL} \\\textbf{(ns)}\end{tabular} & \begin{tabular}[c]{@{}c@{}}\textbf{Power} \\\textbf{($\mu$W)}\end{tabular} & \begin{tabular}[c]{@{}c@{}} \textbf{Energy}\\ \textbf{(pJ)} \end{tabular} & \begin{tabular}[c]{@{}c@{}}\textbf{Area} \\\textbf{($\mu$\textbf{m\textsuperscript{2}})}\end{tabular} & \begin{tabular}[c]{@{}c@{}}\textbf{CPL} \\\textbf{(ns)}\end{tabular} & \begin{tabular}[c]{@{}c@{}}\textbf{Power} \\\textbf{($\mu$W)}\end{tabular} & \begin{tabular}[c]{@{}c@{}} \textbf{Energy}\\ \textbf{(pJ)} \end{tabular} & \begin{tabular}[c]{@{}c@{}}\textbf{Area} \\\textbf{($\mu$\textbf{m\textsuperscript{2}})}\end{tabular} & \begin{tabular}[c]{@{}c@{}}\textbf{CPL} \\\textbf{(ns)}\end{tabular} & \begin{tabular}[c]{@{}c@{}}\textbf{Power} \\\textbf{($\mu$W)}\end{tabular} & \begin{tabular}[c]{@{}c@{}} \textbf{Energy}\\ \textbf{(pJ)} \end{tabular} & \begin{tabular}[c]{@{}c@{}}\textbf{Area} \\\textbf{($\mu$\textbf{m\textsuperscript{2}})}\end{tabular} & \begin{tabular}[c]{@{}c@{}}\textbf{CPL} \\\textbf{(ns)}\end{tabular} & \begin{tabular}[c]{@{}c@{}}\textbf{Power} \\\textbf{($\mu$W)}\end{tabular} & \begin{tabular}[c]{@{}c@{}} \textbf{Energy}\\ \textbf{(pJ)} \end{tabular} \\ 
\hline
\begin{tabular}[c]{@{}c@{}}$\sin(x)$\\ $\tanh(x)$\\ $\arctan(x)$\textsuperscript{\ding{64}} \end{tabular} & \begin{tabular}[c]{@{}c@{}}256 \\512 \\1024\end{tabular} & \begin{tabular}[c]{@{}c@{}}\textbf{435} \\\textbf{497} \\\textbf{554}\end{tabular} & \begin{tabular}[c]{@{}c@{}}0.42 \\0.43 \\0.44\end{tabular} & \begin{tabular}[c]{@{}c@{}}\textbf{706.5} \\\textbf{759.7} \\\textbf{812.8}\end{tabular} & \begin{tabular}[c]{@{}c@{}} \textbf{76.0}\\ \textbf{167.3}\\ \textbf{366.2}
 \end{tabular} & \begin{tabular}[c]{@{}c@{}}\textbf{430} \\\textbf{487} \\\textbf{544}\end{tabular} & \begin{tabular}[c]{@{}c@{}}0.42 \\0.43 \\0.44\end{tabular} & \begin{tabular}[c]{@{}c@{}}\textbf{666.0} \\\textbf{740.6} \\\textbf{777.9}\end{tabular} & \begin{tabular}[c]{@{}c@{}} \textbf{71.6}\\ \textbf{163.0}\\ \textbf{350.5}
 \end{tabular} & \begin{tabular}[c]{@{}c@{}}655 \\734 \\801\end{tabular} & \begin{tabular}[c]{@{}c@{}}0.39 \\0.40 \\0.42\end{tabular} & \begin{tabular}[c]{@{}c@{}}1965.3 \\2092.7 \\2178.2\end{tabular} & \begin{tabular}[c]{@{}c@{}} 196.2\\ 428.6\\ 936.8
 \end{tabular} & \begin{tabular}[c]{@{}c@{}}618 \\683 \\757\end{tabular} & \begin{tabular}[c]{@{}c@{}}0.39 \\0.40 \\0.41\end{tabular} & \begin{tabular}[c]{@{}c@{}}1786.0 \\1914.5 \\2042.6\end{tabular} & \begin{tabular}[c]{@{}c@{}} 178.3\\ 392.1\\ 857.6
 \end{tabular} \\ 
\hline
$\cos(x)$ & \begin{tabular}[c]{@{}c@{}}256 \\512 \\1024\end{tabular} & \begin{tabular}[c]{@{}c@{}}\textbf{489} \\\textbf{575} \\\textbf{632}\end{tabular} & \begin{tabular}[c]{@{}c@{}}0.44 \\0.46 \\0.44\end{tabular} & \begin{tabular}[c]{@{}c@{}}\textbf{761.7} \\\textbf{767.7} \\\textbf{829.8}\end{tabular} & \begin{tabular}[c]{@{}c@{}} \textbf{85.8}\\ \textbf{180.8}\\ \textbf{390.9}
 \end{tabular} & \begin{tabular}[c]{@{}c@{}}\textbf{479} \\\textbf{565} \\\textbf{621}\end{tabular} & \begin{tabular}[c]{@{}c@{}}0.45 \\0.46 \\0.46\end{tabular} & \begin{tabular}[c]{@{}c@{}}\textbf{705.5} \\\textbf{704.2} \\\textbf{780.5}\end{tabular} & \begin{tabular}[c]{@{}c@{}} \textbf{81.3}\\ \textbf{165.8}\\ \textbf{367.6}
 \end{tabular} & \begin{tabular}[c]{@{}c@{}}723 \\797 \\883\end{tabular} & \begin{tabular}[c]{@{}c@{}}0.38 \\0.39 \\0.42\end{tabular} & \begin{tabular}[c]{@{}c@{}}2159.3 \\2296.3 \\2284.0\end{tabular} & \begin{tabular}[c]{@{}c@{}} 210.0\\ 458.5\\ 982.3
 \end{tabular} & \begin{tabular}[c]{@{}c@{}}692 \\766 \\852\end{tabular} & \begin{tabular}[c]{@{}c@{}}0.41 \\0.41 \\0.42\end{tabular} & \begin{tabular}[c]{@{}c@{}}1887.5 \\2041.6 \\2180.8\end{tabular} & \begin{tabular}[c]{@{}c@{}} 198.1\\ 402.3\\ 937.9
 \end{tabular} \\ 
\hline
$\tan(x)$ & \begin{tabular}[c]{@{}c@{}}256 \\512 \\1024\end{tabular} & \begin{tabular}[c]{@{}c@{}}\textbf{1286} \\\textbf{1513} \\\textbf{1630}\end{tabular} & \begin{tabular}[c]{@{}c@{}}0.54 \\0.54 \\0.55\end{tabular} & \begin{tabular}[c]{@{}c@{}}\textbf{1826.7} \\\textbf{1925.7} \\\textbf{1994.8}\end{tabular} & \begin{tabular}[c]{@{}c@{}} \textbf{252.5}\\ \textbf{532.4}\\ \textbf{1123.5}
 \end{tabular} & \begin{tabular}[c]{@{}c@{}}\textbf{1273} \\\textbf{1493} \\\textbf{1609}\end{tabular} & \begin{tabular}[c]{@{}c@{}}0.53 \\0.55 \\0.53\end{tabular} & \begin{tabular}[c]{@{}c@{}}\textbf{1783.6} \\\textbf{1860.8} \\\textbf{2040.9}\end{tabular} & \begin{tabular}[c]{@{}c@{}} \textbf{242.0}\\ \textbf{524.0}\\ \textbf{1107.6}
 \end{tabular} & \begin{tabular}[c]{@{}c@{}}1728 \\1924 \\2092\end{tabular} & \begin{tabular}[c]{@{}c@{}}0.49 \\0.49 \\0.49\end{tabular} & \begin{tabular}[c]{@{}c@{}}3600.4 \\4178.0 \\4473.5\end{tabular} & \begin{tabular}[c]{@{}c@{}} 451.6\\ 1048.2\\ 2244.6
 \end{tabular} & \begin{tabular}[c]{@{}c@{}}1653 \\1850 \\2018\end{tabular} & \begin{tabular}[c]{@{}c@{}}0.48 \\0.45 \\0.47\end{tabular} & \begin{tabular}[c]{@{}c@{}}3594.7 \\4326.1 \\4540.0\end{tabular} & \begin{tabular}[c]{@{}c@{}} 441.7\\ 996.7\\ 2185.0
 \end{tabular} \\ 
\hline
$sigmoid(x)$ & \begin{tabular}[c]{@{}c@{}}256 \\512 \\1024\end{tabular} & \begin{tabular}[c]{@{}c@{}}\textbf{490} \\\textbf{567} \\\textbf{628}\end{tabular} & \begin{tabular}[c]{@{}c@{}}0.44 \\0.45 \\0.47\end{tabular} & \begin{tabular}[c]{@{}c@{}}\textbf{~739.8} \\\textbf{~835.1} \\\textbf{858.9}\end{tabular} & \begin{tabular}[c]{@{}c@{}} \textbf{83.3}\\ \textbf{192.4}\\ \textbf{413.4}
 \end{tabular} & \begin{tabular}[c]{@{}c@{}}\textbf{480} \\\textbf{557} \\\textbf{618}\end{tabular} & \begin{tabular}[c]{@{}c@{}}0.45 \\0.45 \\0.46\end{tabular} & \begin{tabular}[c]{@{}c@{}}\textbf{671.0} \\\textbf{787.0} \\\textbf{830.0}\end{tabular} & \begin{tabular}[c]{@{}c@{}} \textbf{90.7}\\ \textbf{181.3}\\ \textbf{391.0}
 \end{tabular} & \begin{tabular}[c]{@{}c@{}}707 \\785 \\852\end{tabular} & \begin{tabular}[c]{@{}c@{}}0.39 \\0.40 \\0.41\end{tabular} & \begin{tabular}[c]{@{}c@{}}1969.6 \\2172.9 \\2264.3\end{tabular} & \begin{tabular}[c]{@{}c@{}} 222.5\\ 463.7\\ 989.0
 \end{tabular} & \begin{tabular}[c]{@{}c@{}}687 \\764 \\831\end{tabular} & \begin{tabular}[c]{@{}c@{}}0.38 \\0.38 \\0.40\end{tabular} & \begin{tabular}[c]{@{}c@{}}1921.4 \\2267.3 \\2355.8\end{tabular} & \begin{tabular}[c]{@{}c@{}} 196.4\\ 463.6\\ 989.0
 \end{tabular} \\ 
\hline
$e$\textsuperscript{$-x$} & \begin{tabular}[c]{@{}c@{}}256 \\512 \\1024\end{tabular} & \begin{tabular}[c]{@{}c@{}}\textbf{481} \\\textbf{537} \\\textbf{602}\end{tabular} & \begin{tabular}[c]{@{}c@{}}0.43 \\0.50 \\0.46\end{tabular} & \begin{tabular}[c]{@{}c@{}}\textbf{~653.7} \\\textbf{~613.0} \\\textbf{734.9}\end{tabular} & \begin{tabular}[c]{@{}c@{}} \textbf{72.0}\\ \textbf{156.9}\\ \textbf{346.2}
 \end{tabular} & \begin{tabular}[c]{@{}c@{}}\textbf{481} \\\textbf{537} \\\textbf{602}\end{tabular} & \begin{tabular}[c]{@{}c@{}}0.43 \\0.50 \\0.46\end{tabular} & \begin{tabular}[c]{@{}c@{}}\textbf{653.7} \\\textbf{613.0} \\\textbf{734.9}\end{tabular} & \begin{tabular}[c]{@{}c@{}} \textbf{72.0}\\ \textbf{156.9}\\ \textbf{346.2}
 \end{tabular} & \begin{tabular}[c]{@{}c@{}}701 \\778 \\845\end{tabular} & \begin{tabular}[c]{@{}c@{}}0.37 \\0.39 \\0.40\end{tabular} & \begin{tabular}[c]{@{}c@{}}2064.0 \\2195.5 \\2302.2\end{tabular} & \begin{tabular}[c]{@{}c@{}} 195.5\\ 438.9\\ 943.0
 \end{tabular} & \begin{tabular}[c]{@{}c@{}}701 \\778 \\845\end{tabular} & \begin{tabular}[c]{@{}c@{}}0.38 \\0.39 \\0.40\end{tabular} & \begin{tabular}[c]{@{}c@{}}1981.7 \\2198.0 \\2399.9\end{tabular} & \begin{tabular}[c]{@{}c@{}} 192.8\\ 438.9\\ 983.0
 \end{tabular} \\ 
\hline
$Sinc(x)$ & \begin{tabular}[c]{@{}c@{}}256 \\512 \\1024\end{tabular} & \begin{tabular}[c]{@{}c@{}}\textbf{432} \\\textbf{494} \\\textbf{554}\end{tabular} & \begin{tabular}[c]{@{}c@{}}0.41 \\0.42 \\0.43\end{tabular} & \begin{tabular}[c]{@{}c@{}}\textbf{742.6} \\\textbf{764.8} \\\textbf{809.7}\end{tabular} & \begin{tabular}[c]{@{}c@{}} \textbf{78.0}\\ \textbf{164.5}\\ \textbf{338.0}
 \end{tabular} & \begin{tabular}[c]{@{}c@{}}\textbf{422} \\\textbf{483} \\\textbf{544}\end{tabular} & \begin{tabular}[c]{@{}c@{}}0.41 \\0.42 \\0.43\end{tabular} & \begin{tabular}[c]{@{}c@{}}\textbf{682.2} \\\textbf{712.8} \\\textbf{767.7}\end{tabular} & \begin{tabular}[c]{@{}c@{}} \textbf{71.6}\\ \textbf{153.3}\\ \textbf{338.0}
 \end{tabular} & \begin{tabular}[c]{@{}c@{}}619 \\698 \\757\end{tabular} & \begin{tabular}[c]{@{}c@{}}0.38 \\0.40 \\0.39\end{tabular} & \begin{tabular}[c]{@{}c@{}}1900.2 \\1978.8 \\2288.4\end{tabular} & \begin{tabular}[c]{@{}c@{}} 184.9\\ 405.3\\ 958.0
 \end{tabular} & \begin{tabular}[c]{@{}c@{}}598 \\678 \\736\end{tabular} & \begin{tabular}[c]{@{}c@{}}0.37 \\0.37 \\0.39\end{tabular} & \begin{tabular}[c]{@{}c@{}}1842.5 \\2039.5 \\2154.1\end{tabular} & \begin{tabular}[c]{@{}c@{}} 174.5\\ 407.2\\ 860.3
 \end{tabular} \\ 
\hline
$\ln(1+x)$ & \begin{tabular}[c]{@{}c@{}}256 \\512 \\1024\end{tabular} & \begin{tabular}[c]{@{}c@{}}\textbf{475} \\\textbf{551} \\\textbf{608}\end{tabular} & \begin{tabular}[c]{@{}c@{}}0.45 \\0.45 \\0.45\end{tabular} & \begin{tabular}[c]{@{}c@{}}\textbf{620.6} \\\textbf{736.0} \\\textbf{744.6}\end{tabular} & \begin{tabular}[c]{@{}c@{}} \textbf{71.5}\\ \textbf{169.6}\\ \textbf{343.1}
 \end{tabular} & \begin{tabular}[c]{@{}c@{}}\textbf{475} \\\textbf{551} \\\textbf{608}\end{tabular} & \begin{tabular}[c]{@{}c@{}}0.45 \\0.45 \\0.45\end{tabular} & \begin{tabular}[c]{@{}c@{}}\textbf{620.6} \\\textbf{736.0} \\\textbf{744.6}\end{tabular} & \begin{tabular}[c]{@{}c@{}} \textbf{71.5}\\ \textbf{169.6}\\ \textbf{343.1}
 \end{tabular} & \begin{tabular}[c]{@{}c@{}}699 \\779 \\846\end{tabular} & \begin{tabular}[c]{@{}c@{}}0.38 \\0.41 \\0.40\end{tabular} & \begin{tabular}[c]{@{}c@{}}1938.5 \\2105.2 \\2373.4\end{tabular} & \begin{tabular}[c]{@{}c@{}} 193\\ 450.6\\ 972.1
 \end{tabular} & \begin{tabular}[c]{@{}c@{}}701 \\779 \\846\end{tabular} & \begin{tabular}[c]{@{}c@{}}0.38 \\0.39 \\0.40\end{tabular} & \begin{tabular}[c]{@{}c@{}}1983.3 \\2256.5 \\2332.0\end{tabular} & \begin{tabular}[c]{@{}c@{}} 193\\ 450.6\\ 955.2
 \end{tabular} \\
\hline
\end{tabular}
\justify{\scriptsize{ \underline{\textbf{\ding{93} and \ding{70} are the modification of Works in~\cite{Trig-Parhi} and \cite{9444648}, respectively}}. \ding{64}: The hardware design of these functions are the same; The coefficients are different. Power consumption is reported at the maximum frequency. 
}}
\label{hw_results}
\vspace{-1em}
\end{table*}

Parhi and Liu~\cite{Trig-Parhi} 
investigated 
SC designs for implementing complex transcendental 
functions using polynomial approximations, where 
positive and negative coefficients were arranged in decreasing order for each polynomial function. 
Chu et al.~\cite{9444648} proposed a 
correlation-based SC implementation of polynomial functions with unipolar bit-streams. 
They introduced a \texttt{NAND}-\texttt{AND} structure with fewer delay elements to realize trigonometric functions. Both approaches rely on Horner's rule for 
polynomial evaluation. 

Similarly, our proposed \Design~applies Horner's rule to 
the Maclaurin series expansion of the target function.
In existing SOTA designs, 
two separate RNG units are used to generate independent bit-streams for inputs and 
coefficients. All coefficient bit-streams are generated from a shared RNG unit~\cite{Trig-Parhi,9444648}. 
To decorrelate the intermediate bit-streams, these works employ additional delay elements (e.g. \texttt{D-FFs}). 
In contrast, \Design~\textit{eliminates the need for such delay elements}
by leveraging the LD and distinct distribution properties of 
VDC-$2^n$ sequences. 
This allows both input and coefficient bit-streams to be generated using a single 
VDC-$2^n$ BSG unit, resulting in significant hardware cost reductions while maintaining accuracy and independence in intermediate computations.

Fig.~\ref{designs} presents the set of transcendental functions implemented within the proposed \Design~framework. 
We revisit the CLB of prior SOTA  designs by integrating our newly developed BSG inputs. 
In the revised designs, the delay blocks (previously required to decorrelate bit-streams) are either completely eliminated  or significantly reduced. 
Figs.~\ref{designs}(a) and \ref{designs}(b) illustrate the enhanced designs for $\mathbf{sin(x)}$ and $\mathbf{cos(x)}$ functions, respectively. 
The key distinctions between our design and the SOTA approaches are twofold: i) the use of a versatile, single-source quasi-random BSG unit based on VDC-$2^n$ sequences, and ii) the removal of the mid-stage decorrelation blocks (e.g., \texttt{D-FF}s), which reduces both  hardware cost and computational latency. 

For $\mathbf{tan(x)}$, shown in Fig.~\ref{designs}(c), we utilize an efficient SC division circuit that derives results from $\mathbf{sin(x)}$ and $\mathbf{cos(x)}$. This circuit integrates a correlator block (consisting of a down counter and an \texttt{AND} gate) and the CORDIV architecture~\cite{7560183}. 
Figs.~\ref{designs}(d)-(i) depict the~\Design~design of additional transcendental functions, including $\mathbf{tanh(x)}$, $\mathbf{arctan(x)}$, $\mathbf{sigmoid(x)}$, $\mathbf{Sinc(x)}$, $\mathbf{e\textsuperscript{\textbf{$-x$}}}$, and $\mathbf{ln(1+x)}$.

\section{Experimental Results}
\label{experiments}
The VDC-$2^n$-based BSG inherently exhibits quasi-randomness, providing the desired statistical independence between mid-level circuit elements without requiring additional decorrelation components.
Our evaluation demonstrates that  \Design~achieves significantly higher accuracy compared to SOTA designs. Tables~\ref{acc_results} and \ref{acc_results2} present a comparative accuracy analysis between \Design-based and SOTA designs. 

For the SOTA baselines, we used maximal period LFSRs (two LFSRs defined by the polynomials $x^{10}+x^{8}+x^{6}+1$ and $x^{10}+x^{8}+x^{5}+x^{4}$) and executed each operation $1,000$ times to ensure statistically significant results. Additionally, we tested the SOTA designs with Sobol sequences (generated using \texttt{MATLAB}'s built-in Sobol sequence generator), applied separately to the input and coefficient BSGs. However, incorporating these quasi-random sequences into SOTA designs degraded performance, as delay elements disrupted the independence properties of the sequences. 
Our findings indicate that when quasi-random RNGs are used in SC implementations of transcendental functions, delay elements must be minimized or eliminated to preserve independence. 
Owing to their substantial hardware cost, Sobol sequences were not adopted in \Design.

The input BSGs in the \Design~framework are parameterized 
using different 
VDC-$2^n$ configurations, as detailed in Tables~\ref{acc_results} and \ref{acc_results2}. 
The coefficient BSG module generates bit-streams representing the function-specific coefficients for each transcendental function. 
For each test case, we evaluate all possible input values within the $[0, N]$ interval, 
corresponding to a 
$\log_2N$-bit precision. We explore 
multiple VDC-$2^n$ sequences for both input and coefficient bit-streams and 
report the mean squared error (MSE) along with
the total number of decorrelating delay elements used in any stage of the circuit. 

As shown, the \Design~designs eliminate all mid-level delay elements,  unlike prior approaches.
A key advantage of our methodology is its ability to maintain high
accuracy even with significantly shorter bit-stream lengths. 
This is clearly demonstrated in Tables~\ref{acc_results} and \ref{acc_results2}, where \Design~outperforms SOTA designs even when the bit-stream length is reduced to as low as $N=64$.
Furthermore, while existing designs require two separate BSGs (e.g., LFSR1/LFSR2 or Sobol-1/Sobol-2,\linebreak as shown in Tables~\ref{acc_results} and \ref{acc_results2}), \Design~achieves independence with a single BSG unit 
eliminating the need for additional hardware. 
For example, in \Design-based design of $\sin(x)$ function, 
VDC-128, VDC-256, and VDC-512 sequences are assigned to the coefficients $\frac{1}{42}$, $\frac{1}{20}$, and $\frac{1}{6}$, respectively, when using $N=1024, 512$ (first and second rows of Table~\ref{acc_results} for $\sin(x)$ function). 

We compare the hardware costs of the proposed \Design~framework against SOTA designs for implementing various transcendental 
functions. 
We synthesized all designs using Synopsys Design Compiler v2018.06 with the 45$nm$ FreePDK gate library.
Table~\ref{hw_results} summarizes the synthesis results in terms of 
area, critical path latency (CPL), power consumption at the maximum working frequency, and energy consumption for each design. 
As shown, \Design~consistently outperforms the SOTA methods by utilizing a highly efficient BSG as the core of the SC architecture. By eliminating mid-stage delay elements and consolidating multiple BSGs into a single unit, \Design~achieves significant hardware efficiency gains. 
Specifically, the proposed designs reduce hardware area, power, and energy consumption by up to 33\%, 72\%, and 64\%, respectively, while maintaining computational accuracy.

\begin{figure}[b]
\vspace{-1.5em}
  \centering
  \includegraphics[width=\linewidth]{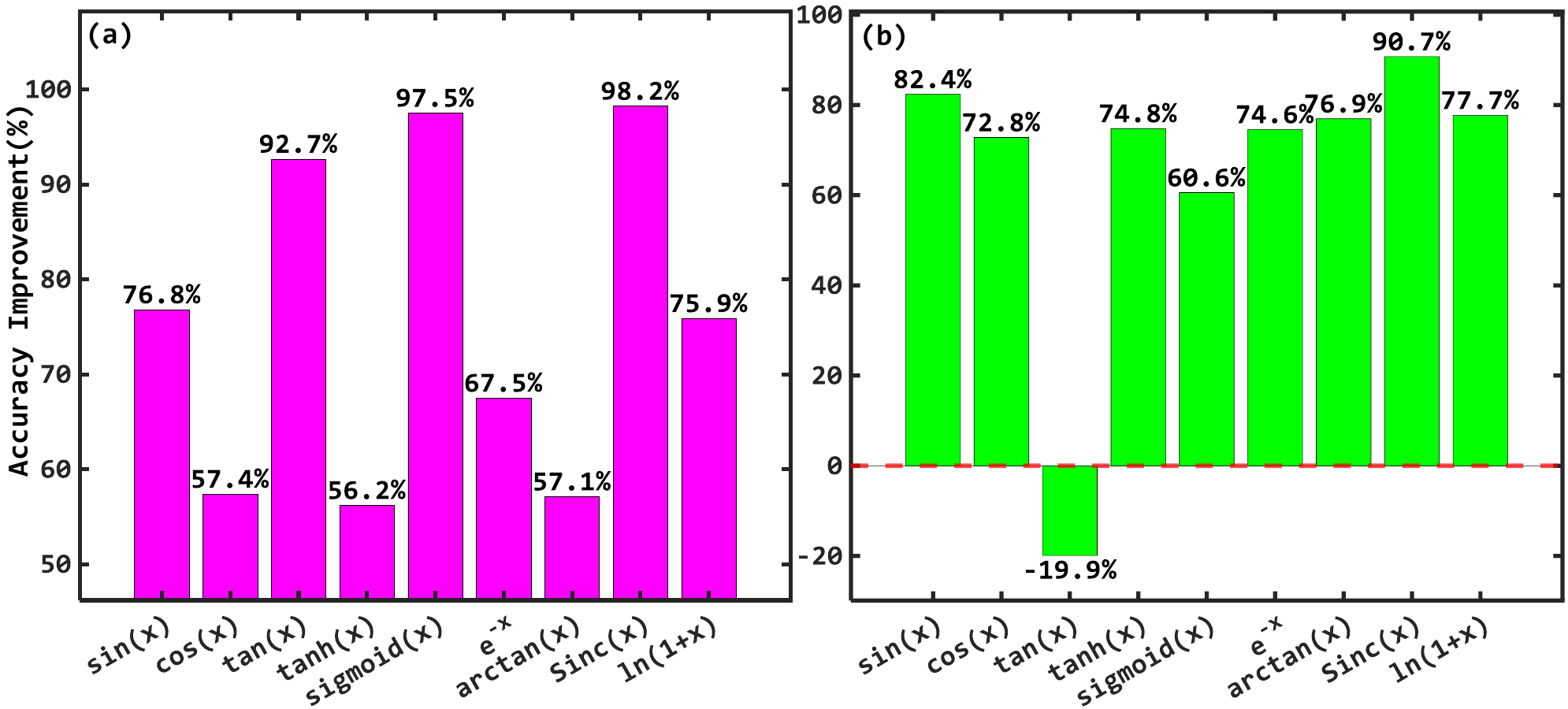}
  \vspace{-2em}
  \color{black}
  \caption{An overview of accuracy improvements: (a) \Design~over the work in~\cite{Trig-Parhi} and (b) \Design~over the work in~\cite{9444648}.}  
  \label{performance_plots}
  
\end{figure}

\begin{figure}[t]
  \centering
  \includegraphics[width=\linewidth]{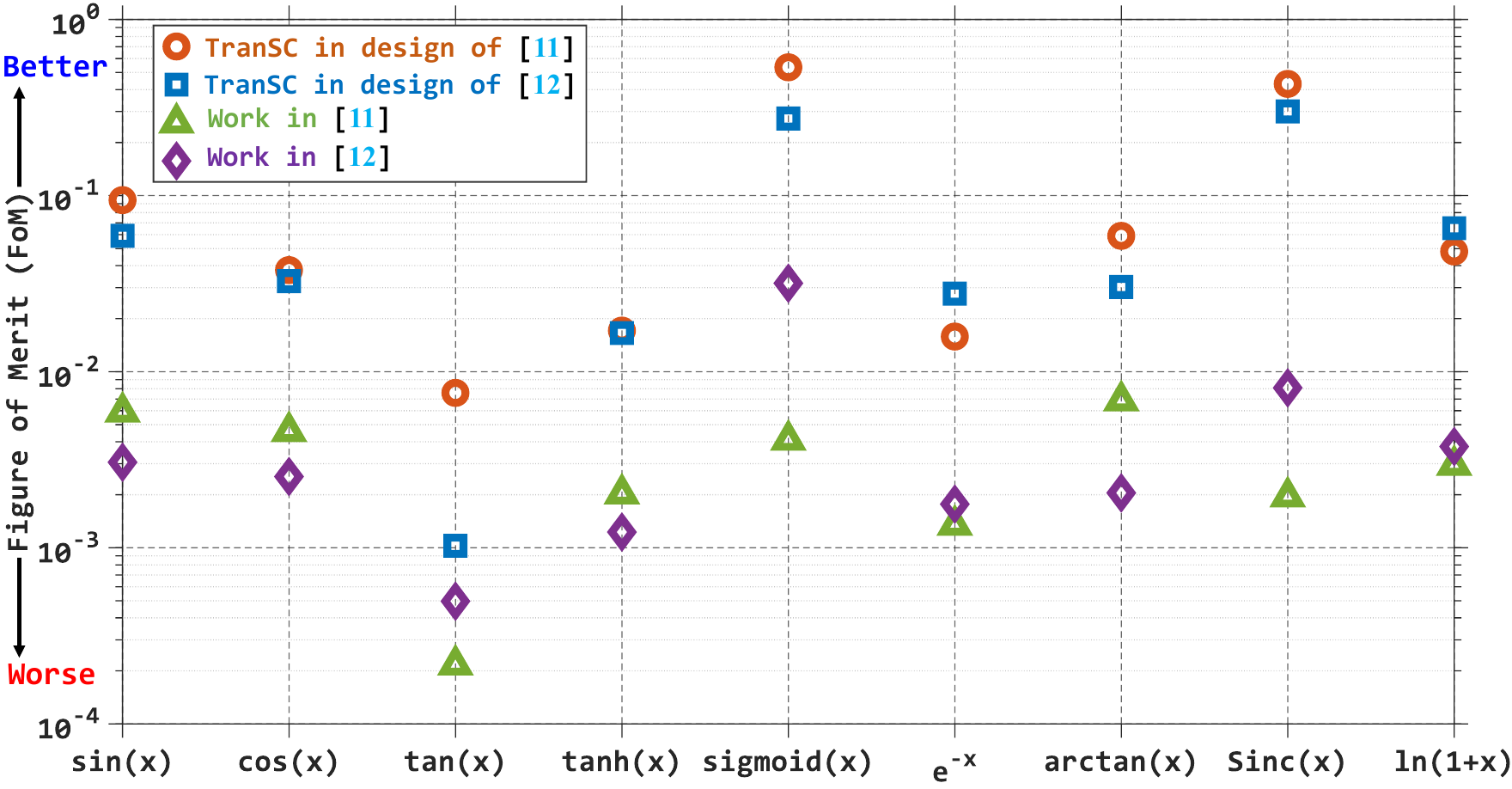}
  \vspace{-1.5em}
  \color{black}
  \caption{Comparison of \textit{Figure of Merit} metric for different design approaches.}  
  \label{fom}
  \vspace{-1.em}
\end{figure}

\vspace{-0.5em}
\section{Proof of Concept}
\label{proof_concept}
\subsection{Performance Analysis}

By incorporating distinct and independent BSGs, the proposed \Design~framework eliminates the need for decorrelator blocks (i.e., delay elements) in  intermediate stages, thereby enhancing hardware efficiency compared to SOTA designs. 
Fig.~\ref{performance_plots} illustrates the accuracy improvements (y-axis) achieved by \Design~over SOTA implementations across various transcendental functions (x-axis), using a fixed bit-stream length of $N=1024$.

To further validate the effectiveness of the 
\Design~design, we evaluate the \textit{Figure of Merit (FoM)}, a comprehensive metric that combines accuracy and hardware (HW) costs to reflect overall design efficiency. The \textit{FoM} is defined as:
\vspace{-0.5em}
\begin{equation}
    \label{FoM_eq}
    FoM = \frac{Accuracy}{HW Costs}
        = \frac{\frac{1}{MSE}}{Area\times Power\times Latency}
\end{equation}
\vspace{-.75em}

\noindent This metric provides a unified basis for comparing different designs by jointly considering MSE, hardware footprint, power consumption, and latency.
Fig.~\ref{fom} presents the \textit{FoM} evaluation across various designs, using data extracted from Tables~\ref{acc_results}, \ref{acc_results2}, and \ref{hw_results}. As defined in Equation~\ref{FoM_eq}, a higher \textit{FoM} value indicates superior overall performance, achieved through a combination of higher computational accuracy and reduced hardware cost.
\vspace{-1.em}


\color{black}
\subsection{Why does VDC-$2^n$ perform well?}
To further explain why the \Design~designs--powered by 
VDC-$2^n$ sequences--outperform the SOTA 
methods; we evaluated 
the $SCC$ (Equation~\ref{SCC_equation}) and the \textit{Zero Correlation Error ($ZCE$)} metrics. $ZCE$ quantifies the degree of independence between SC bit-streams, producing a value of zero for configurations that exhibit maximal independence~\cite{ZCE_ASPDAC21}. It is formally defined as:
\begin{equation}
\label{ZCE_equation}
ZCE = \Delta \cdot \left( 1 - \frac{\Delta_0}{\Delta} \right)
\end{equation}
where, $\Delta = \frac{a}{N} - \frac{(a + b)(a + c)}{N^2}$, and $\Delta_0 = \frac{\left\lfloor \frac{(a + b)(a + c)}{N} + \frac{1}{2} \right\rfloor}{N} - \frac{(a + b)(a + c)}{N^2}$. Similar to the $SCC$ metric, the values $a$, $b$, $c$, and $d$ in the $ZCE$ formulation represent the occurrence counts of logic pairs $11$, $10$, $01$, and $00$, respectively, obtained from corresponding bit positions of two bit-streams. Using both $SCC$ and $ZCE$ metrics, we evaluate the design performance through a representative case: the $\mathbf{\sin(x)}$ function.
Figs.~\ref{difference}(a) and (b) compare the conventional method~\cite{Trig-Parhi} with the proposed \Design~design, highlighting the intermediate computational terms for each approach as follows:

{\small{
\begin{equation}
    \begin{aligned}
       i_1&=X^2 \\ 
       i_2&=1-\frac{1}{42}i_1 = 1-\frac{1}{42}X^2 \\
       i_3&=1-\frac{1}{20}i_1i_2 = 1-\frac{1}{20}X^2(1-\frac{1}{42}X^2) \\
       i_4&=1-\frac{1}{6}i_1i_3 = 1-\frac{1}{6}X^2(1-\frac{1}{20}X^2(1-\frac{1}{42}X^2)).\\
    \end{aligned}
\end{equation}
}
}

\begin{figure}[t]
  \centering
  \includegraphics[width=\linewidth]{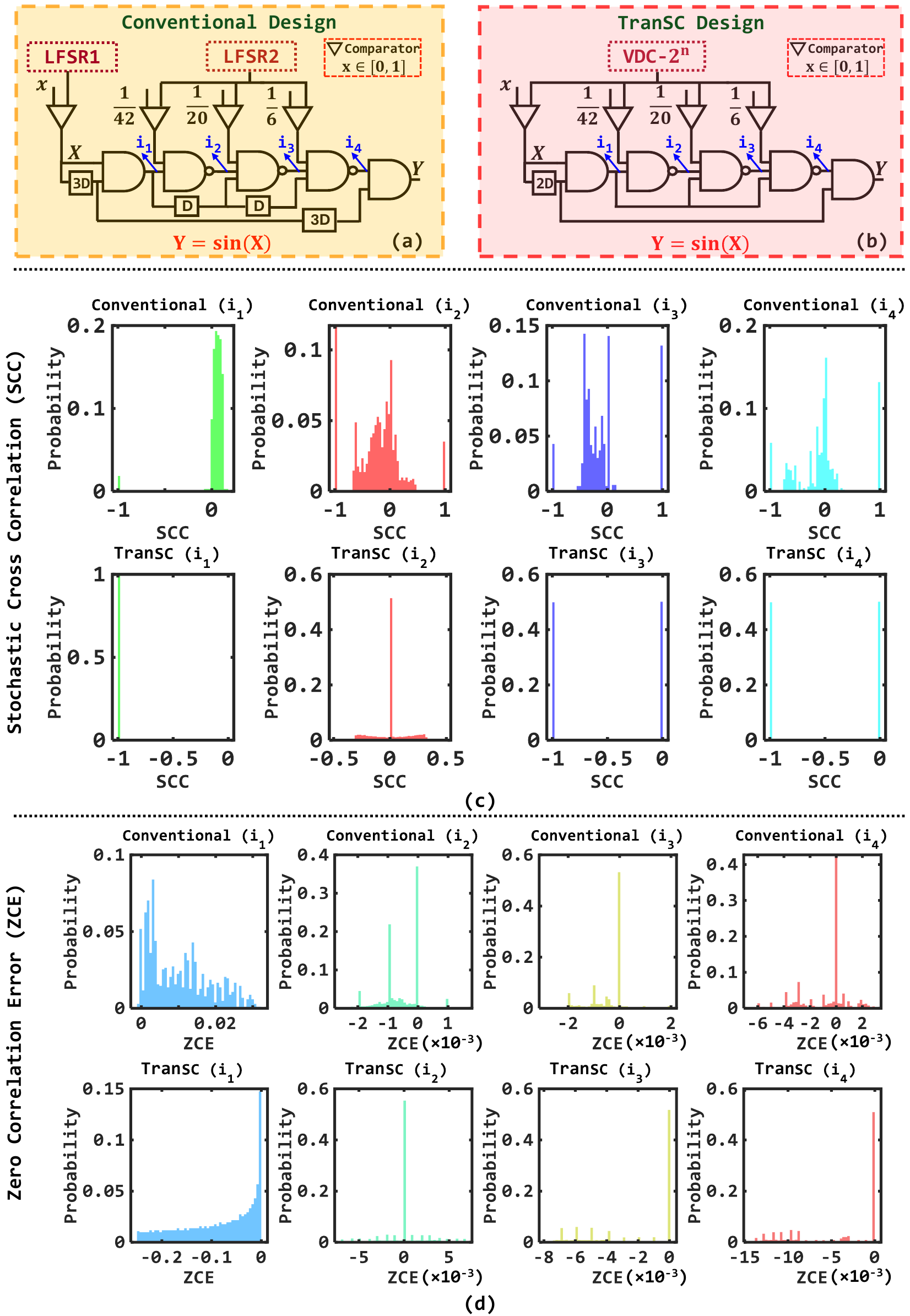}
  \vspace{-2.25em}
  \color{black}
  \caption{Measuring the independence level of intermediate bit-streams in $\mathbf{sin(x)}$ function. (a) Conventional design~\cite{Trig-Parhi}, (b) \Design~design, (c) $SCC$ measurement, and (d) ZCE measurement considering conventional and \Design~approaches for intermediate bit-streams of $i_1$, $i_2$, $i_3$, and $i_4$.}  
  \label{difference}
  \vspace{-1.5em}
\end{figure}

\color{black}

\begin{figure*}[t]
  \centering
  \includegraphics[width=0.9\linewidth]{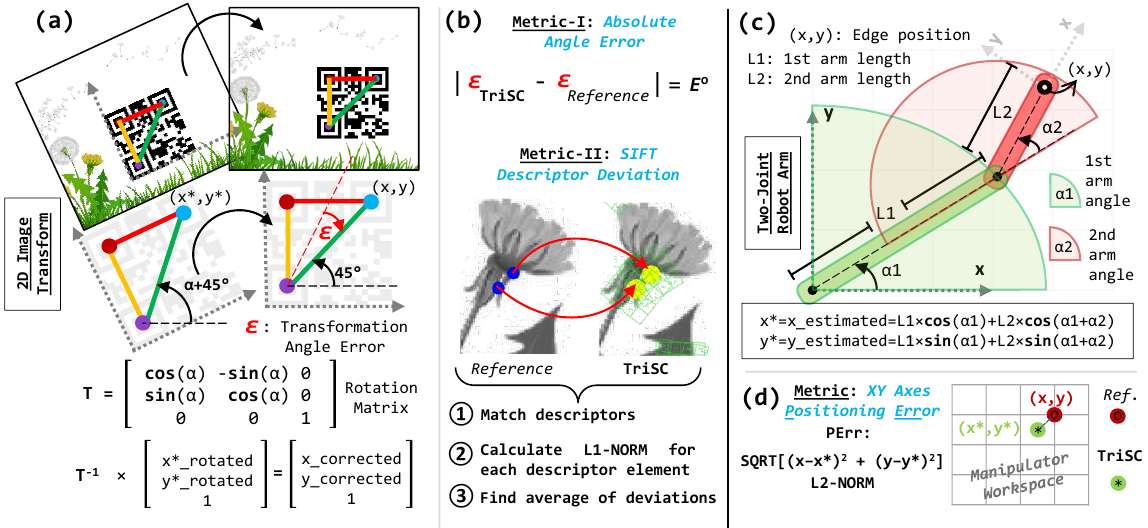}
  \vspace{-1em}
  \caption{
  \Design~use cases: (a) 
  2D image transformation with visual markers, (b) Absolute angle error ($E^{\circ}$) and SIFT descriptor deviation (SIFT$_{\sigma}$), (c) 
  Two-joint arm kinematics, (d) Manipulator position estimation error (\textit{PErr}).}
  \label{applications}
  \vspace{-1.em}
\end{figure*}

Fig.~\ref{difference}(c) presents the $SCC$ measurements--expressed in terms of normalized probability--for each intermediate bit-stream. To ensure fair comparison and consistency with the  SOTA designs, we employed SC bit-streams of length $1024$, adhering to the design configurations specified in Table~\ref{acc_results}. 
These measurements were obtained by evaluating all possible 10-bit precision input values within the $[0,1]$ interval. 

In the first intermediate stage ($i_1$), the conventional (SOTA) method exhibits at most $20\%$ zero correlation ($SCC=0$), whereas the \Design~design consistently achieves minimum correlation ($SCC=-1$) across all bit-streams. 
In the second intermediate stage ($i_2$), the $SCC$ values for the conventional design are broadly distributed within the range of $-1$ to $+1$, indicating 
considerable dependency among bit-streams. 
Conversely, the \Design~design shows $SCC$ values tightly clustered around $0$, with minimal spread, demonstrating 
superior decorrelation between bit-streams in this stage. 
For the third and fourth intermediate stages ($i_3,i_4$), the conventional design continues to show a wide $SCC$ distribution--some clustering near $0$, but with significant fluctuations throughout the full range of ($-1$ to $+1$). Meanwhile, the \Design~design maintains 
either perfect independence ($SCC=0$) or minimum correlation ($SCC=-1$) for the $i_3$ and $i_4$ stages.
Notably, the $SCC$ trends observed in the \Design~design align with those of 
the conventional approach only when the latter explicitly uses decorrelator elements to enforce independence. 
The fluctuation patterns in the $SCC$ plots of the SOTA design are attributed to the inherent randomness and correlation drift in LFSR-based generators. In contrast, the \Design~framework--leveraging the LD nature of VDC sequences--avoids such inconsistencies and delivers stable inter-stream correlation properties.

Fig.~\ref{difference}(d) illustrates the $ZCE$ plots as an alternative metric for evaluating the independence 
between SC bit-streams in the intermediate stages of the $\mathbf{\sin(x)}$ function designs. 
In the first intermediate stage ($i_1$), the conventional design exhibits $ZCE$ values that peak near $0$ but show 
a noticeable spread towards higher positive errors. In contrast, the \Design~design shows $ZCE$ values tightly clustered around $0$ with minimal variation, indicating stronger independence. 
In the second stage ($i_2$), the $ZCE$ values for the conventional design span a wide range, with peaks near $0$ but substantial dispersion extending toward both positive and negative extremes. Meanwhile,  the \Design~design maintains $ZCE$ values closely centered around $0$, again with minimal deviation. 
For the third stage ($i_3$), the conventional design exhibits an even broader  $ZCE$ distribution, covering a wide spectrum of values. In contrast, the \Design~design retains a narrow $ZCE$ distribution concentrated near $0$, signifying better inter-stream decorrelation.
Finally, in the fourth stage ($i_4$), the conventional design exhibits considerable variability in $ZCE$ values, including peaks near $0$ along with significant outliers in both directions. In contrast, the \Design~design continues to show $ZCE$ values densely concentrated around $0$, with very limited deviation, highlighting its ability to maintain inter-stream independence consistently across all computation stages.

These analyses collectively validate the effectiveness of the proposed \Design~design, establishing it as a lightweight yet highly accurate solution for efficiently implementing transcendental functions with the SC paradigm. Building on this foundation, we next demonstrate the practical applicability of \Design~in real-world scenarios--introducing SC-based designs, for the first time, to the domains of 
image transformation and robotics.


\vspace{-1em}
\section{Use-Case Studies}
\label{design}
In this section, we evaluate the performance of 
the proposed \Design~design approach in two novel applications, explored 
for the first time in the SC literature: \ding{192}~\textit{2D Image Transformation} and \ding{193} \textit{Robot Arm Kinematics}. 

\textit{2D Image Transformation} is a fundamental operation widely employed in  Quick Response (QR) code-based image rectification, with many applications in mobile robot navigation and camera posture calibration. 
These tasks often require 
computationally intensive trigonometric operations, demanding 
low-cost hardware implementations~\cite{PEDERSEN2016282}. 
Fig.~\ref{applications}(a) illustrates the square finder patterns of a QR code in a still image, which are used to determine the reference rotation angle.
A typical QR code features three finder patterns located at the top-left, top-right, and bottom-left corners.
The diagonal line connecting these patterns (depicted in green) forms a 45$^{\circ}  $ angle with the $x$-axis;  any deviation $\alpha$ from this ideal angle signifies a rotation of the image.
To correct for such rotation, a transformation matrix $T$ is applied, which encapsulates common image manipulation operations such as translation, scaling, shear, reflection, and rotation. 
In particular, the rotation operations in $T$ involve \textit{trigonometric} functions such as \textit{sine} and \textit{cosine}. The inverse transformation $\frac{1}{T}$ is crucial in rectifying image orientation by compensating for any rotation angle $\alpha$. 
Conventional QR code processing also employs the inverse of $T$ to reverse any rotation incurred during image acquisition, thereby restoring the image to its correct alignment. 
Fig.~\ref{applications}(a) illustrates the rotation-wise $T$ matrix, which employs 
\textit{sine} and \textit{cosine} operations. 
The inverse operation is used to determine the corrected pixel positions 
$x\_corrected$ and $y\_corrected$.

Fig.~\ref{applications}(b) presents the evaluation metrics we use 
 to assess the 
 performance. 
\textit{Absolute Angle Error} assesses the performance of the SC designs (proposed and SOTA) 
compared to the reference binary model, based on the transformation angle error $\epsilon$. Following image correction, $\epsilon$ indicates the deviation error from the expected angle, as depicted in Fig.~\ref{applications}(a). 
\textit{SIFT Descriptor Deviation} evaluates the deviation from scale-invariant feature transform (SIFT)-based descriptors. We compare the output 
images 
with reference images after transformation, and analyze the matched descriptors 
using the $\mathcal{L}1$ norm to quantify average alterations when using 
the implemented designs 
in image transformation.

We also employ the SC-based \textit{trigonometric} designs 
in a \textit{robotic kinematics} application. 
Fig.~\ref{applications}(c) illustrates the operations 
for a 2-joint robotic arm involving a manipulator system with 
two links with lengths of $L1$ and $L2$. 
The 
$\alpha_1$ and $\alpha_2$ angles define the movement ranges of links shown in green and red shaded regions in the $xy$-coordinate. The manipulator edge point $( x , y )$ is 
estimated using the forward kinematics equations that involve 
\textit{sine}
and \textit{cosine} 
functions. 
When SC operations 
are used for calculating 
edge positioning, we compare the estimated points $  ( x^* , y^* )  $ with the binary calculation as the reference to find the positioning error (\textit{PErr}) in Fig.~\ref{applications}(d). 

\begin{table}[t]
\centering
\caption{Performance of \Design~in Different Use Cases}
\vspace{-1em}
\label{useCasesResults}

 \setlength\dashlinedash{0.2pt}
 \setlength\dashlinegap{1.5pt}
 \setlength\arrayrulewidth{0.3pt}
 \setlength{\tabcolsep}{0.7pt} 
 \renewcommand{\arraystretch}{1.5}
 \relscale{0.78}
\begin{tabular}{|c|c|c|c|c|c|c|} 
\hline
\multirow{3}{*}{\begin{tabular}[c]{@{}c@{}}\textbf{Design} \\ \textbf{Approach}\end{tabular}} & \multicolumn{3}{c|}{\textbf{Use-Case I: Image Orientation}} & \multicolumn{3}{c|}{\textbf{Use-Case II: Robotics Positioning}} \\ 
\cdashline{2-7}[1pt/1pt]
 & \multicolumn{3}{c|}{\begin{tabular}[c]{@{}c@{}}\textbf{Angle Error ($E^{\circ}$) \&~}\textbf{SIFT Deviation (SIFT$_{\sigma}$)~}\textbf{}\end{tabular}} & \multicolumn{3}{c|}{\begin{tabular}[c]{@{}c@{}}\textbf{Position Estimation Error (PErr)} \textbf{}\end{tabular}} \\ 
\cline{2-7}
 & \textbf{N=256} & \textbf{N=512} & \textbf{N=1024} & \textbf{\ \ N=256 \ \ } & \textbf{N=512} & \textbf{N=1024} \\ 
\hline
\textbf{\Design\textsuperscript{\ding{93}}} & \begin{tabular}[c]{@{}c@{}}$E^{\circ}$: \textbf{0.091}\\SIFT$_{\sigma}$: \textbf{6.368}\end{tabular} & \begin{tabular}[c]{@{}c@{}}$E^{\circ}$: \textbf{0.084} \\SIFT$_{\sigma}$: \textbf{6.367}\end{tabular} & \begin{tabular}[c]{@{}c@{}}$E^{\circ}$: \textbf{0.068} \\SIFT$_{\sigma}$: \textbf{6.366}\end{tabular} & \begin{tabular}[c]{@{}c@{}}

PErr: \textbf{0.158}\end{tabular} & \begin{tabular}[c]{@{}c@{}}PErr: \textbf{0.157}\end{tabular} & \begin{tabular}[c]{@{}c@{}}PErr: \textbf{0.155}\end{tabular} \\ 
\hdashline[1pt/1pt]

\textbf{\Design\textsuperscript{\ding{70}}} & \begin{tabular}[c]{@{}c@{}}$E^{\circ}$: \textbf{0.177}\\SIFT$_{\sigma}$: \textbf{6.457}\end{tabular} & \begin{tabular}[c]{@{}c@{}}$E^{\circ}$: \textbf{0.164}\\SIFT$_{\sigma}$: \textbf{6.405}\end{tabular} & \begin{tabular}[c]{@{}c@{}}$E^{\circ}$: \textbf{0.163}\\SIFT$_{\sigma}$: \textbf{6.373}\end{tabular} & \begin{tabular}[c]{@{}c@{}}

PErr: \textbf{0.176}\end{tabular} & \begin{tabular}[c]{@{}c@{}}PErr: \textbf{0.174}\end{tabular} & \begin{tabular}[c]{@{}c@{}}PErr: \textbf{0.173}\end{tabular} \\
\hdashline[1pt/1pt]

\textbf{Work in~\cite{Trig-Parhi}} & \begin{tabular}[c]{@{}c@{}}$E^{\circ}$: 0.386\\SIFT$_{\sigma}$: 6.594\end{tabular} & \begin{tabular}[c]{@{}c@{}}$E^{\circ}$: 0.207\\SIFT$_{\sigma}$: 6.531\end{tabular} & \begin{tabular}[c]{@{}c@{}}$E^{\circ}$: 0.178\\SIFT$_{\sigma}$: 6.448\end{tabular} & \begin{tabular}[c]{@{}c@{}}

PErr: 0.546\end{tabular} & \begin{tabular}[c]{@{}c@{}}PErr: 0.542\end{tabular} & \begin{tabular}[c]{@{}c@{}}PErr: 0.541\end{tabular} \\ 
\hdashline[1pt/1pt]

\textbf{Work in~\cite{9444648}} & \begin{tabular}[c]{@{}c@{}}$E^{\circ}$: 0.858\\SIFT$_{\sigma}$: 6.602\end{tabular} & \begin{tabular}[c]{@{}c@{}}$E^{\circ}$: 0.652\\SIFT$_{\sigma}$: 6.597\end{tabular} & \begin{tabular}[c]{@{}c@{}}$E^{\circ}$: 0.493\\SIFT$_{\sigma}$: 6.543\end{tabular} & \begin{tabular}[c]{@{}c@{}}

PErr: 0.756\end{tabular} & \begin{tabular}[c]{@{}c@{}}PErr: 0.636\end{tabular} & \begin{tabular}[c]{@{}c@{}}PErr: 0.468\end{tabular} \\
\hline
\end{tabular}
\vspace{-.75em}
\justify{\scriptsize{\ding{93}:  \Design~applied to the design of~\cite{Trig-Parhi}, \ding{70}: \Design~applied to the design of~\cite{9444648}. The best-performing BSGs from Table~\ref{acc_results} are used 
for \Design\textsuperscript{\ding{93}} and \Design\textsuperscript{\ding{70}}. 
}}
 \vspace{-0.5em}
\end{table}

Table~\ref{useCasesResults} displays the 
performance results for the two implemented use cases. 
For the first case, we examine a dataset of QR code images~\cite{10.1145/2448531.2448548} 
captured in a real-world environment and report mean angle error and SIFT descriptor deviations. For the second use case, we analyze navigation ranges of links 
for all possible Cartesian pairs of $( \alpha_1 , \alpha_2 )$ and report the mean \textit{PErr}. 
As it can be seen, across all test scenarios, our proposed architectures outperform
the SOTA designs for different bit-stream lengths from 256 to 1024.

\section{Discussion}
\label{discussion}
\subsection{Comparison Against SOTA RNGs}
Sobol sequences are well known for their excellent uniformity properties and strong performance for SC designs. Our argument for employing VDC-$2^n$ sequences is not that they are universally more uniform, but rather that they are better suited to the architecture requirements of the \Design~framework. The primary objective of \Design~is to eliminate the mid-stage decorrelation elements (\texttt{D-FF}s), which have been a major source of hardware cost and latency in prior SOTA designs. We eliminate these elements 
by generating multiple, distinct, and statistically independent bit-streams for both the function inputs and polynomial coefficients from a \textit{single, shared hardware source}.
The hardware architecture of the VDC-$2^n$ generator—a simple up-counter with configurable bit-reversal hardwiring—is exceptionally lightweight. 
It can produce up to $log_2 N$ distinct and independent sequences simultaneously with virtually zero additional hardware cost. 
In contrast, a Sobol sequence generator is far more complex, 
requiring an address generator, storage for direction vectors, and \texttt{XOR} logic networks~\cite{8327916,Najafi_TVLSI_2019}. Generating multiple independent Sobol streams would necessitate either replicating this complex logic or employing even more complex multi-dimensional generators, both of which substantially increase hardware costs. By comparison, VDC-$2^n$ sequences provide the \Design~framework with a highly efficient BSG unit at a fraction of the hardware cost of Sobol-based alternatives.

The results presented in Tables~\ref{acc_results} and \ref{acc_results2} demonstrate that the \Design~design achieves superior accuracy and stronger decorrelation, as reflected by lower SCC and ZCE values in Fig.~\ref{difference}. These results validate that the uniformity provided by the VDC-$2^n$ sequences is sufficient to ensure the required level of independence in the implemented designs.
Table~\ref{hw_cost_RNG} compares the hardware cost of bit-stream generation across three approaches: an LFSR-based generator, an SOTA FSM-based Sobol generator, and a VDC-$2^n$-based bit-stream generator.
The results show that the VDC-based generator 
achieves about 28\% and 14\% smaller area, while consuming 24\% and 43\% less power than the FSM-based Sobol~\cite{Najafi_FSM_DATE2021} and the LFSR-based bit-stream generator, respectively.
This hardware advantage becomes even more significant when multiple independent sequences are required, thereby justifying the choice of 
VDC-$2^n$ as the core enabler of the lightweight \Design~architecture.

\begin{table}[t]
\centering
\caption{Hardware implementation cost of SC bit-stream generation considering 8-bit precision RNG sources}
\vspace{-1em}
\setlength{\tabcolsep}{11pt}
\begin{tabular}{cccc} 
\toprule
\textbf{RNG Type} & \begin{tabular}[c]{@{}c@{}}\textbf{Area} \\($\mu m^2$)\end{tabular} & \begin{tabular}[c]{@{}c@{}}\textbf{Power@Max Freq.} \\($\mu W$)\end{tabular} & \begin{tabular}[c]{@{}c@{}}\textbf{CPL} \\($ns$)\end{tabular} \\ 
\midrule
LFSR & 233 & 949.2 & 0.37 \\ 
Sobol-FSM~\cite{Najafi_FSM_DATE2021} & 281 & 720.4 & 0.36 \\ 
VDC-$2^n$ & 201 & 544.4 & 0.34 \\
\bottomrule
\end{tabular}
\label{hw_cost_RNG}
\vspace{-0.5em}
\justify{\scriptsize{
The maximal period with the polynomial $x^{8}+x^{6}+x^{3}+x^2$ is used for LFSR.
}}
\vspace{-0.5em}
\end{table}

\vspace{-1.em}
\subsection{Comparison Against Conventional Binary Designs}
\subsubsection{\Design~vs. CORDIC}
The CORDIC (Coordinate Rotation Digital Computer) algorithm~\cite{CORDIC_TCASI_2009} is a classic and widely-used 
``shift-and-add'' technique for computing transcendental functions, particularly efficient on hardware without dedicated multipliers, such as FPGAs or simple microcontrollers. As an iterative method, CORDIC's latency scales with the required precision. 
\Design's accuracy and latency are determined by $N$, while its hardware implementation comprises a BSG and simple combinational logic. The primary distinction lies in their target applications: CORDIC is well-suited for systems that demand high, guaranteed accuracy where iterative latency is acceptable. In contrast, \Design~is optimized for scenarios where ultra-low power, minimal hardware area, and high fault tolerance are the dominant design requirements, and where the inherent inaccuracy 
of SC is tolerable. 
The presented use cases highlight domains where this trade-off proves especially advantageous.

\subsubsection{\Design~vs. Piecewise Polynomial Interpolation (PPI)}
PPI methods 
represent another major class of function evaluators. In these techniques, the input range is partitioned into smaller segments, and a low-degree polynomial, often stored as coefficients in a look-up table, is used to approximate the function within each segment. This approach can achieve high accuracy, often surpassing what is practical with SC-based methods, by increasing either the number of segments or the polynomial degree~\cite{Lee_Interpolation_TC2008}.
A comparison with \Design~highlights a fundamental trade-off in hardware design. 
PPI methods rely on traditional binary arithmetic units (e.g., multipliers and adders) along with memory blocks, with hardware cost scaling with the required precision and polynomial complexity. By contrast, \Design's hardware is dominated by the BSG and simple bit-wise logic, offering a lightweight alternative.

Table~\ref{hw_cost_designs} benchmarks \Design~against CORDIC and a second-order PPI 
method for the $\sin(x)$ function, with all designs targeting a 10-bit equivalent input precision.
While CORDIC and PPI offer significantly higher accuracy (lower MSE), this comes at a substantial cost in hardware area and power consumption. \Design~is nearly $12\times$ and $8\times$ smaller and consumes $18\times$ and $7\times$ less power than CORDIC and PPI designs, respectively. Its only drawback is energy consumption, a known limitation of SC designs due to the serial processing of bit-streams. 
However, this inefficiency can be substantially mitigated through parallelism.
Overall, this quantitative comparison positions \Design~not as a universal replacement for high-accuracy conventional binary methods, but as a competitive solution for a specific class of applications--those that are severely resource-constrained and can tolerate modest accuracy trade-offs in exchange for significant gains in efficiency and robustness.

\begin{table}[t]
\centering
\caption{Cross-Paradigm Performance and Resource Benchmark for $\sin(x)$ Function with 10-bit Binary Precision}
\vspace{-1em}
\resizebox{\columnwidth}{!}{
\begin{tabular}{cccccc} 
\toprule
\textbf{Methodology} & \begin{tabular}[c]{@{}c@{}}\textbf{Area} \\($\mu m^2$)\end{tabular} & \begin{tabular}[c]{@{}c@{}}\textbf{Power@Max} \\ \textbf{Freq.}($\mu W$)\end{tabular} & \begin{tabular}[c]{@{}c@{}}\textbf{CPL} \\($ns$)\end{tabular} & \begin{tabular}[c]{@{}c@{}} \textbf{Energy}\\ (\textbf{$pJ$}) \end{tabular} & \begin{tabular}[c]{@{}c@{}} \textbf{MSE}\\ ($\times 10^{-4}$) \end{tabular} \\ 
\midrule
CORDIC & 6562 & 14500 & 0.60 & 8.7 & $\ll 0.01$\\ 
PPI & 4419 & 5500 & 0.90 & 4.9 & $\ll 0.005$ \\ 
\Design & \textbf{\underline{554}}$\downarrow$ & \textbf{\underline{812.8}}$\downarrow$ & \textbf{\underline{0.44}}$\downarrow$ & 366.2$\uparrow$ & 0.523$\uparrow$ \\
\bottomrule
\end{tabular}
}
\label{hw_cost_designs}
\vspace{-0.6em}
\justify{\scriptsize{
CORDIC uses unrolled 10-iteration hardware implementation. PPI uses ROMs and multiplier-accumulator units designed for 
2nd-order polynomials.
}}
\vspace{-0.5em}
\end{table}

\color{black}

\section{Conclusions}
\label{conclusion}
This work introduced a hardware-efficient methodology for implementing transcendental 
functions using SC with quasi-random Van der Corput (VDC) sequences. The proposed \Design~framework eliminates the need for decorrelator elements, thereby reducing both hardware cost and latency while achieving higher accuracy compared to state-of-the-art SC designs. We demonstrated the effectiveness of the approach through extensive evaluation across various functions, including two novel SC applications in robotics vision and maneuvering. The results show the potential of \Design~for lightweight and efficient systems that rely on  trigonometric and transcendental computations. Future work will explore the development of end-to-end SC architectures, particularly in neural networks, by incorporating our designs for non-linear functions such as sigmoid activations and logarithmic functions. 
These advancements could further extend the applicability of SC to emerging domains such as robotic learning and low-power AI systems.

\color{black}

\vspace{-0.5em}
\bibliographystyle{IEEEtran}
\bibliography{bibliography,Hassan}


\begin{IEEEbiography}[{\includegraphics[width=1.3in,height=1.25in,clip,keepaspectratio]{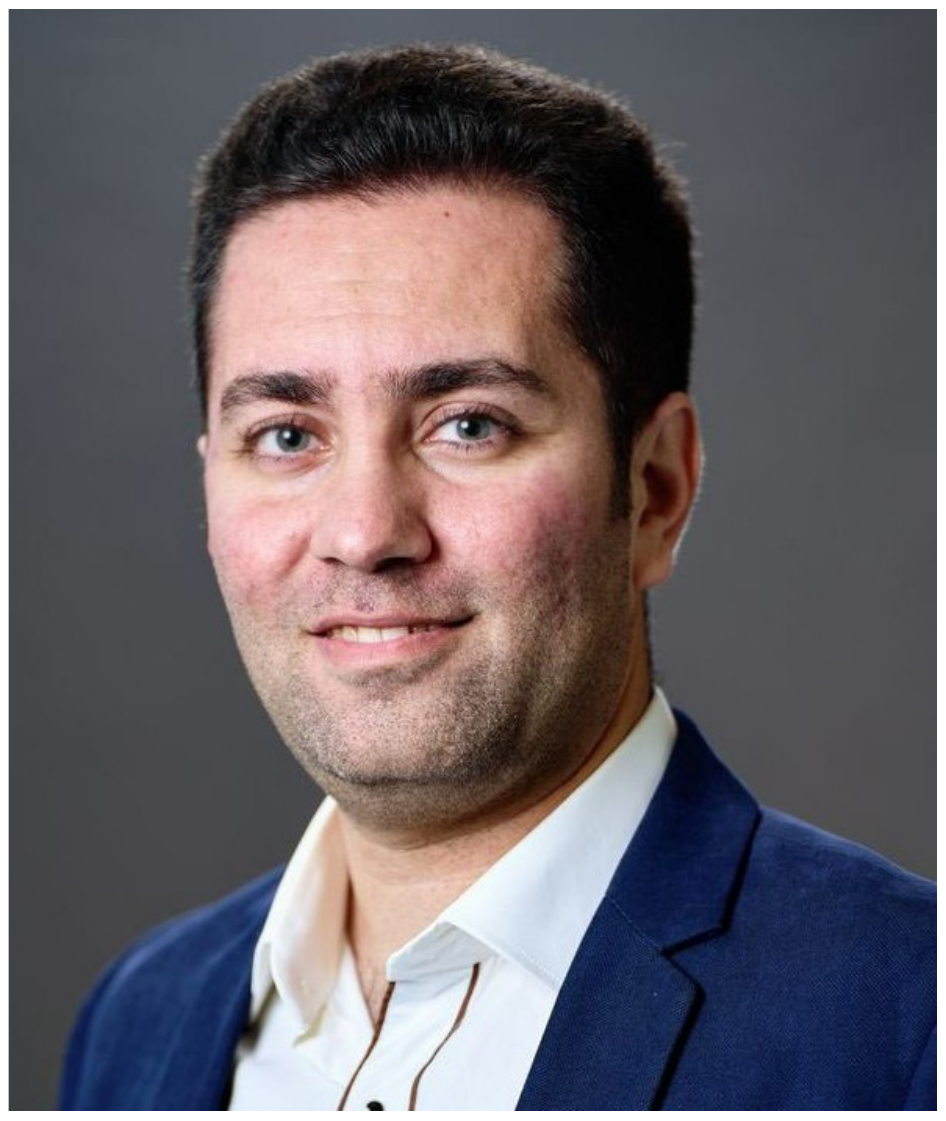}}]{Mehran Moghadam} (S’22) received his B.Sc. degree in Computer Engineering, and M.Sc. degree in \textit{Computer Systems Architecture} from the University of Isfahan, Iran, in 2010 and 2016. He graduated as one of the top-ranking students in both programs.
In 2022, he began his Ph.D. studies in Computer Engineering at the School of Computing and Informatics, University of Louisiana at Lafayette, Lafayette, LA, USA. In 2024, he transferred to the Electrical, Computer, and Systems Engineering department at Case Western Reserve University, Cleveland, OH, USA, to continue pursuing his Ph.D. in Computer Engineering.
Mehran became a finalist in the ACM SIGBED Student Research Competition (SRC) at ESWEEK and ICCAD in 2024 and was selected as a DAC Young Fellow in DAC 2024. His research interests include emerging and unconventional computing paradigms, such as energy-efficient stochastic computing models, real-time and high-accuracy hyperdimensional computing systems, bit-stream processing, robust in-memory arithmetic computation, and low-power in/near-sensor computing designs for edge AI. 
\end{IEEEbiography}

\vspace{-3.em}


\begin{IEEEbiography}[{\includegraphics[width=1.1in,height=1.25in,clip,keepaspectratio]{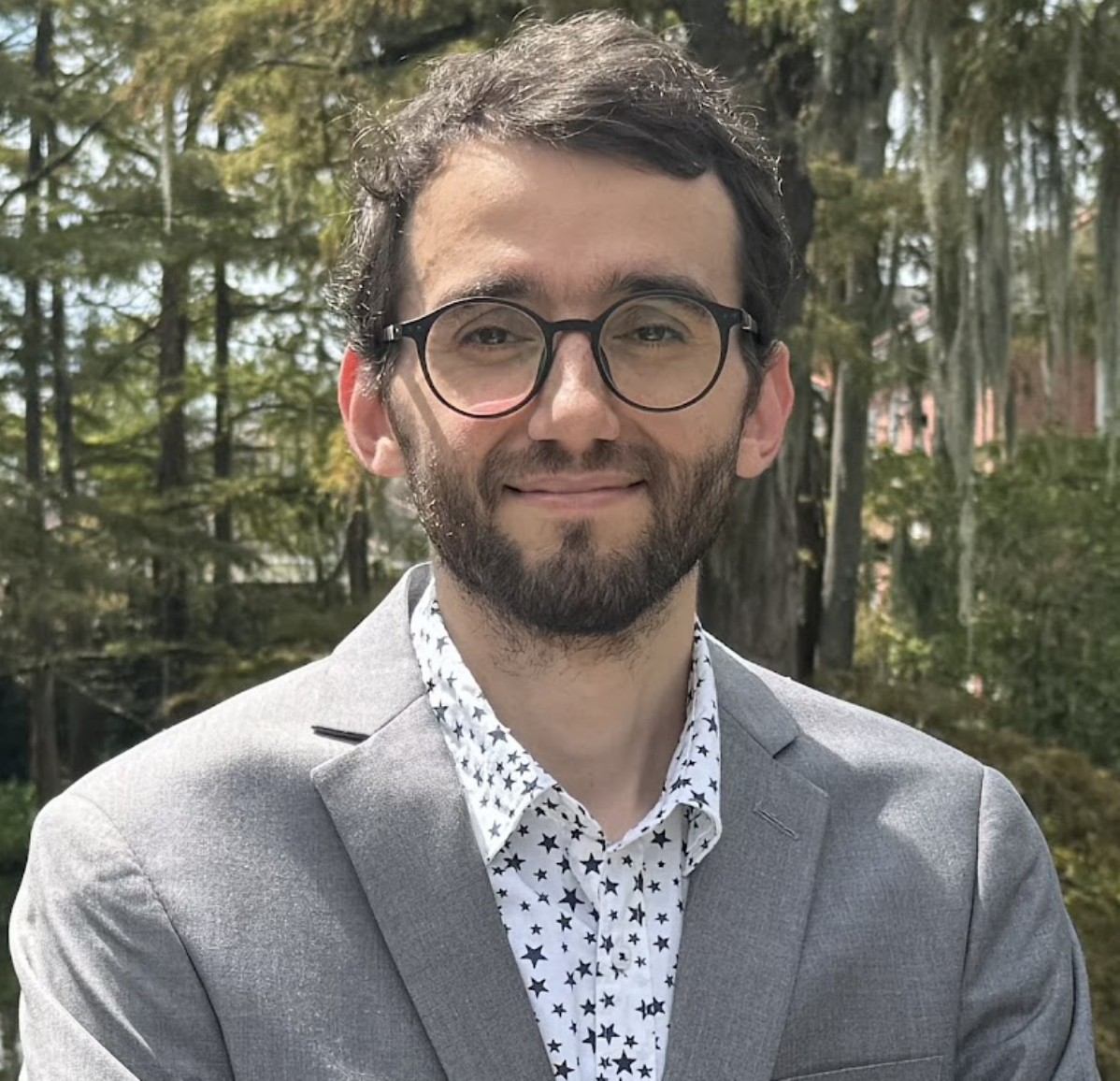}}]{Sercan Aygun} (S’09-M’22-SM'24) received a B.Sc. degree in Electrical \& Electronics Engineering and a double major in Computer Engineering from Eskisehir Osmangazi University, Turkey, in 2013. He completed his M.Sc. degree in Electronics Engineering from Istanbul Technical University in 2015 and a second M.Sc. degree in Computer Engineering from Anadolu University in 2016. Dr. Aygun received his Ph.D. in Electronics Engineering from Istanbul Technical University in 2022. 
Dr. Aygun received the Best Scientific Research Award of the ACM SIGBED Student Research Competition (SRC) ESWEEK 2022, the Best Paper Award at GLSVLSI'23, and the Best Poster Award at GLSVLSI'24. Dr. Aygun's Ph.D. work was recognized with the Best Scientific Application Ph.D. Award by the Turkish Electronic Manufacturers Association and was also ranked first nationwide in the Science and Engineering Ph.D. Thesis Awards by the Turkish Academy of Sciences. He is currently an Assistant Professor at the School of Computing and Informatics, University of Louisiana at Lafayette, Lafayette, LA, USA. He works on tiny machine learning and emerging computing, including stochastic and hyperdimensional computing. \end{IEEEbiography}




\vspace{-2.em}
\begin{IEEEbiography}[{\includegraphics[width=1.3in,height=1.25in,clip,keepaspectratio]{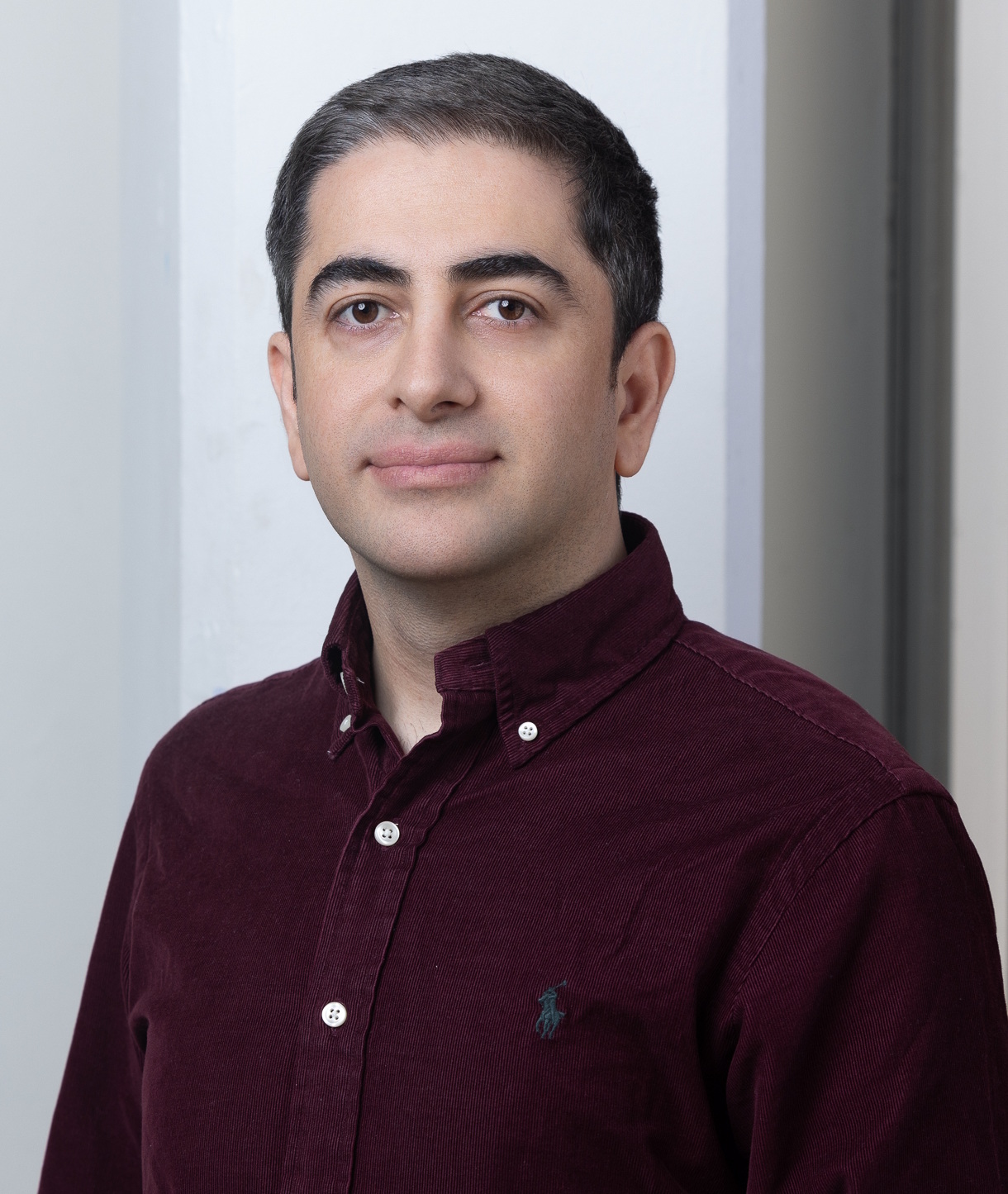}}]{M. Hassan Najafi} (S’15-M’18-SM'23) received the B.Sc. degree in Computer Engineering from the University of Isfahan, Iran, the M.Sc. degree in Computer Architecture from the University of Tehran, Iran, and the Ph.D. degree in Electrical Engineering from University of Minnesota, Twin Cities, USA, in 2011, 2014, and 2018, respectively. He was an Assistant Professor at the School of Computing and Informatics, University of Louisiana at Lafayette, Lafayette, LA, USA, from 2018 to 2024. He is currently an Associate Professor at the Electrical, Computer, and Systems Engineering Department at Case Western Reserve University, Cleveland, OH, USA. His research interests include stochastic and approximate computing, unary processing, in-memory computing, and hyperdimensional computing. He has authored/co-authored more than 80 peer-reviewed papers and has been granted 6 U.S. patents, with more pending. Dr. Najafi received the NSF CAREER Award in 2024, the Best Paper Award at GLSVLSI'23 and ICCD’17, the Best Poster Award at GLSVLSI'24, the 2018 EDAA Outstanding Dissertation Award, and the Doctoral Dissertation Fellowship from the University of Minnesota. Dr. Najafi has been an editor for the IEEE Journal on Emerging and Selected Topics in Circuits and Systems and a Technical Program Committee Member for many EDA conferences. \end{IEEEbiography}

\end{document}